\documentclass[aps,showpacs,PRX,twocolumn,superscriptaddress]{revtex4-2}
\makeatletter

\newcommand{\Rmnum}[1]{\expandafter\@slowromancap\romannumeral #1@}
\makeatother
\usepackage{graphicx}
\usepackage{CJK}
\usepackage{color}
\usepackage[colorlinks,bookmarks=false,citecolor=blue,linkcolor=red,urlcolor=blue]{hyperref}
\usepackage{multirow}
\usepackage{ulem}
\usepackage{tabularx}
\usepackage[nounderscore]{syntax}
\newcommand{\ket}[1]{\mbox{$ | #1 \rangle $}}
\newcommand{\bra}[1]{\mbox{$ \langle #1 | $}}
\newcommand{\tetrahedron}{
    \includegraphics[height=1.3ex]{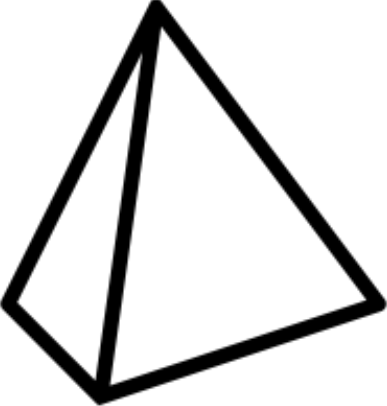} % \scriptstyle
}
\newcommand{\be}{\begin{equation}}
\newcommand{\ee}{\end{equation}}

\usepackage[english]{babel}
\usepackage{amsmath}
\usepackage{graphicx}
\usepackage{subfigure}
\usepackage{color}
\usepackage{amsfonts}
\usepackage{tikz}
\usepackage{esint}
\usepackage{mathrsfs}
\usepackage{verbatim}

\newcommand{\bea}{\begin{eqnarray}}
\newcommand{\eea}{\end{eqnarray}}

\usepackage{dcolumn}% Align table columns on decimal point
\usepackage{bm}% bold math
\usepackage{subfigure} % 子图包
\usepackage{color}

\begin{document}

\preprint{APS/123-QED}

%\title{Doped resonating valence bond states: Robustness of the spin ice phases in 3D Rydberg arrays}
\title{Doped resonating-valence-bond states:
Robustness of the spin-ice phases in three-dimensional Rydberg arrays}

\author{Jingya Wang}
\affiliation{State Key Laboratory of Surface Physics and Department of Physics, Fudan University, Shanghai 200438, China}
\affiliation{Department of Physics, School of Science and Research Center for Industries of the Future, Westlake University, Hangzhou 310030,  China}
\affiliation{Institute of Natural Sciences, Westlake Institute for Advanced Study, Hangzhou 310024, China}

\author{Changle Liu}
\email{liuchangle89@gmail.com}
%\affiliation{Independent Researcher, China}
\affiliation{School of Physics and Mechatronic Engineering, Guizhou Minzu University, Guiyang 550025, China}

\author{Yan-Cheng Wang}
\email{ycwangphys@buaa.edu.cn}
\affiliation{Hangzhou International Innovation Institute, Beihang University, Hangzhou 311115, China}
\affiliation{Tianmushan Laboratory, Hangzhou 311115, China}

\author{Zheng Yan}
\email{zhengyan@westlake.edu.cn}
\affiliation{Department of Physics, School of Science and Research Center for Industries of the Future, Westlake University, Hangzhou 310030,  China}
\affiliation{Institute of Natural Sciences, Westlake Institute for Advanced Study, Hangzhou 310024, China}
\date{\today}

\begin{abstract}
Rydberg blockade effect provides a convenient platform for simulating locally constrained many-body systems, such as quantum dimer models and quantum loop models, especially their novel phases like topological orders and gapless quantum spin ice (QSI) phases. To discuss the possible phase diagram containing different QSIs in three-dimensional (3D) Rydberg arrays, we have constructed an extended Rokhsar-Kivelson (RK) Hamiltonian with equal weight superposition ground state in different fillings at the RK point. Therefore, the perfect QSIs with fixed local dimer filling and their monomer-doped states can be simulated directly by Monte Carlo sampling. Using single-mode approximation, the excitations of dimers and monomers have also been explored in different fillings.
We find that, in the thermodynamical limit, even doping a small amount of monomers can disrupt the topological structure and lead to the existence of off-diagonal long-range order. However, in a finite size (as in cold-atom experiment), the property of QSI will be kept in a certain region like a crossover after doping. The phase diagram containing different QSIs and off-diagonal order phases is proposed.

\end{abstract}

\maketitle

\section{\label{sec:level1}Introduction}
%###################################################
%\cite{mmoessner2021} #update ref-old  [51]
In recent years, programmable Rydberg atom arrays have played the roles of powerful quantum simulators for the investigation of highly correlated and entangled quantum matter~\cite{endres2016atom,bernien2017probing,Browaeys.2020,morgado2021quantum,bluvstein2022quantum,gonzalez2024observation}. These systems have opened up new avenues to study novel quantum many-body phases \cite{de2019observation,Ebadi.2021,scholl2021quantum,Semeghini21,yan2023emergent}, complex dynamics \cite{turner2018weak,keesling2019quantum,bluvstein2021controlling}, gauge theories \cite{Celi1,Surace1,qudit,IGT}, and even combinatorial optimization problems \cite{pichler2018quantum,Ebadi2022,yan2021sweeping}.

An especially promising direction that has recently attracted much attention is the simulation of quantum phases of matter in these tunable atomic setups. Such phases and the transitions between them have been intensely studied for Rydberg atoms arrayed in one spatial dimension \cite{samajdar2018numerical,whitsitt2018quantum,PhysRevLett.122.017205,chepiga2021kibble} as well as in various two-dimensional (2D) geometries, including square \cite{Samajdar_2020,PhysRevLett.126.170603,kalinowski2021bulk,Kim.2021,orourke2022entanglement}, triangular \cite{li2022quantum,zhou2022quantum_simulation}, honeycomb \cite{honeycomb,trimer}, kagome \cite{Samajdar.2021,yan2022triangular}, and ruby lattices~\cite{Verresen.2020, hannes_dynamical,zhai}. 
In particular, Ref.~\cite{Samajdar.2021} identified an intriguing highly correlated regime in the phase diagram of the kagome-lattice Rydberg atom array, characterized by a lack of symmetry-breaking solid order and a large entanglement entropy. 
The correlations in this region were found to be ``liquid-like'', in that the Rydberg excitations are highly correlated and constrained by strong Rydberg-Rydberg repulsions, as opposed to a weakly interacting gas where atomic excitations at different sites fluctuate almost individually. 
Such a correlated liquid phase can be understood if we map this system onto a quantum dimer model (QDM) on the triangular lattice \cite{moessner2011quantum,moessner2001ising,roychowdhury2015z,yan2021topological,ran2024cubic,ZY2022loop}. 
Under this mapping, the Rydberg blockade effect can dynamically exert certain local constraints, such that each dual triangular lattice site is attached to some fixed number of dimers. 
On the non-bipartite triangular lattice, this local constraint further enables mapping to the $\mathbb{Z}_2$ lattice gauge theory,
which raises the possibility of liquid phases with $\mathbb{Z}_2$ topological order \cite{Samajdar.2021,yan2021topological,yan2022triangular}. 
Moreover, the dimer density can be tuned continuously by laser detuning, which acts as an effective chemical potential of dimers.
For the example of the kagome lattice, dimer filling from one per site to two per site has been found in numerics~\cite{Samajdar.2021,yan2022triangular,yan2023emergent}.

On the other hand, Rydberg experiments have also been extended to investigations of QDM physics on bipartite lattices. In contrast to the cases of non-bipartite lattices that host $\mathbb{Z}_2$ gauge structures, QDMs on bipartite lattices generally host a continuous U(1) gauge structure. 
In three-dimensional (3D) bipartite lattices, QDMs \cite{RK88,moessner2011quantum} have long been known to host 
the so-called Coulomb phases that exhibit deconfined charges and gapless ``photon'' excitations, just analogous to Maxwell's theory of electromagnetism~\cite{huse2003coulomb,alet2006unconventional,mossner2003ising,moessner2003three,moessner2021spin}. Such a U(1) Coulomb phase has been extensively studied in the context of pyrochlore lattice frustrated quantum magnets and was named the U(1) quantum spin ice (QSI) in that context.  
In two spatial dimensions, however, the ``photons'' will be gapped out due to the Polyakov's confinement mechanism~\cite{moessner2011quantum,zhou2021emergent,zhou2022quantum,zhou2023quantum,zhou2022quantum_simulation}, hence the deconfined U(1) Coulomb phase cannot exist as a stable phase in 2D. As such, the experimental realization of the Coulomb QSI phase requires Rydberg simulators in three spatial dimensions~\cite{shannon2012quantum,huang2018dynamics}. 

In recent years, with the development of quantum simulation technologies, platforms such as Rydberg atom arrays have provided new avenues for the nonequilibrium preparation of topological quantum states. Notably, the topological features observed on these platforms may not originate from the equilibrium ground state of the system. For instance, Ref.~\cite{Semeghini21} reported spin-liquid–like behavior 
that was in fact realized through nonequilibrium dynamical processes. Subsequent theoretical studies have revealed the underlying mechanism: the response rates of e-anyons and m-anyons to external parameter variations differ significantly—while e-anyons can adiabatically follow the parameter changes and become annihilated, m-anyons are preserved due to dynamical freezing. As a result, a quantum spin liquid with few defects emerges at the macroscopic scale Ref.~\cite{sahay2022quantum}. This mechanism has not only been confirmed by extensive numerical simulations but also exhibits strong universality, thereby opening new directions for the preparation and stabilization of topological states in finite systems.

 The recent Ref.\cite{shah2024quantumspinicethreedimensional}  pointed out that the low-energy effective model of the pyrochlore Rydberg array is the diamond quantum dimer model. They give an experimental scheme to realize the quantum spin ice phases based on this mapping between the dimer model and the Rydberg array. 
The effective model we adopt is highly related to this experimental proposal.

In this work, we explore a crucial yet previously unexplored extension of the QDM on 3D bipartite lattices 
\cite{huse2003coulomb,alet2006unconventional,moessner2006quantum,moessner2011quantum}. 
The main aim of our work is to explore how robust the spin ice in this system, thus the quantum spin ice of Rydberg array can be effectively considered as the resonating valence bond (RVB) in 3D quantum dimer models. Under realistic experimental conditions, the local constraint (Rydberg blockade) cannot be perfectly enforced, and as a result, some monomers usually remain. Therefore, the key question is, do the spin ice still exist after doping in RVB, and are there some remaining characteristics of spin ice in a finite size system even if the spin ice disappear? Our numerical results actually have answered these questions generally, not limited to the detailed Rydberg arrays.

Unlike the original QDMs, our model allows for a variable dimer density in which monomer doping is incorporated. 
For original QDMs, the local constraint that each site is attached to a given number of dimers is crucial for the emergent gauge structure.
However, in Rydberg experiments, monomer defects violating such local constraint can be presented for various reasons~\cite{Semeghini21}.  
In the previous study, we examined the case of triangular lattice QDM~\cite{yan2022triangular}, which relates to the case where atoms are positioned on the sites of the kagome lattice \cite{roychowdhury2015z,Plat2015z2,Samajdar.2021}. The QDM is experimentally accessible and has already been conducted initial investigations in Ref.~\cite{Semeghini21}, where a dimer model on the kagome lattice.
In this work, we turn our attention to 3D bipartite lattices with emergent U(1) gauge structure, and discover the physics with monomer doping present.  

Besides the inevitable presence of monomers in experiments, the robustness of U(1) Coulomb liquid phase in presence of monomers itself is an issue that is worth further numerical explorations. 
Here we show that with monomer doping, the 3D QDM exhibits novel features relevant to ongoing Rydberg-atom experiments. 
More interestingly, the phase diagram of this extended QDM also reveals several distinct U(1) Coulomb liquids with different dimer fillings on each site. 
Note that in early discussions of such U(1) Coulomb liquids in dimer models, 
the different phases are characterized by their background charges, or, the number of dimers attached to each site~\cite{huang2018dynamics,shah2024quantumspinicethreedimensional,isakov2005spin,nisoli2013colloquium}. In the present model, the number of dimers on each site fluctuates among different values; 
nevertheless, the distinction of the different QSIs still survives based on the different local charges of the background~\cite{ross2011quantum}. 
However, in the case of a soft constraint as is relevant here, we will investigate this fate numerically with Monte Carlo simulations.

In light of the situation, here, we overcome this challenge and present large-scale Monte Carlo simulation results on the Rokhsar-Kivelson (RK) like Hamiltonian with tunable doping in diamond and cubic lattice QDMs. 
Surprisingly, while U(1) Coulomb liquids can be stable in 3D, our numerical results for large system sizes -- in both static and dynamical properties -- reveal the \textit{fragility} of U(1) Coulomb phases upon doping. 
Nevertheless, the good news is that, at smaller length scales, the properties of QSIs are preserved under weak doping. 
Considering that current Rydberg atom array platforms are inherently finite-sized systems, the features of the U(1) Coulomb liquids remain  visible at low monomer doping densities. Therefore, we believe that Rydberg atom experiments offer a promising platform for detecting the property of U(1) Coulomb liquids.

%###################################################
\section{diamond lattice}
\label{section:Pyrochlore lattice}
\begin{figure} [h]%tbp]
\includegraphics[width=0.45\textwidth]{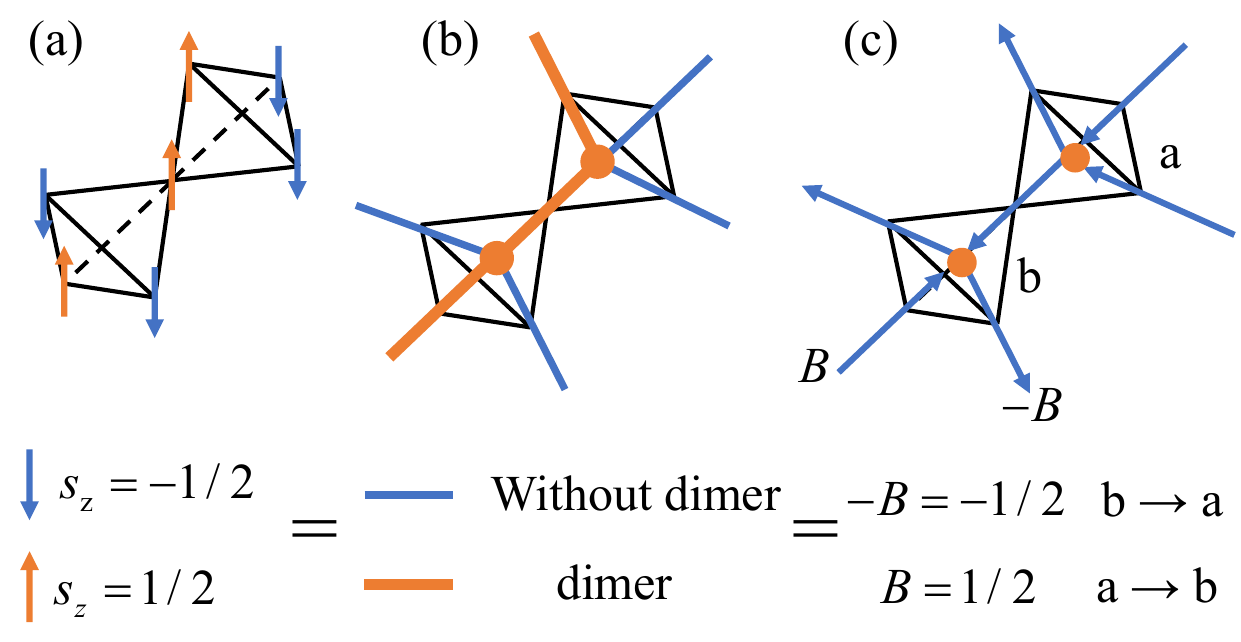}%[scale=0.5]% 
\caption{\label{fig:flux}{The mapping between spin, dimer and flux. (a) Rydberg array can be mapped to the spin model. The atom at the Rydberg excited state is mapped to the spin-up (up arrows), while the atom at the atomic ground state is mapped to the spin-down (down arrows).(b) Spin can be mapped onto dimer (orange line). A dimer, corresponds to a spin pointing upward, $S_z=1/2$, while the absence of a dimer (blue line)} corresponds to a spin pointing downward, $S_z=-1/2$. (c) They can be interpreted as fluxes, we identify a spin pointing up, with a flux vector $B$ from sublattice a to b, and vice versa.}
The ice rule corresponds to the contact of two upward spins, two dimers and the flux field vanishes for each tetrahedron.
\end{figure}

\begin{figure} [h]%tbp]
\includegraphics[width=0.5\textwidth]{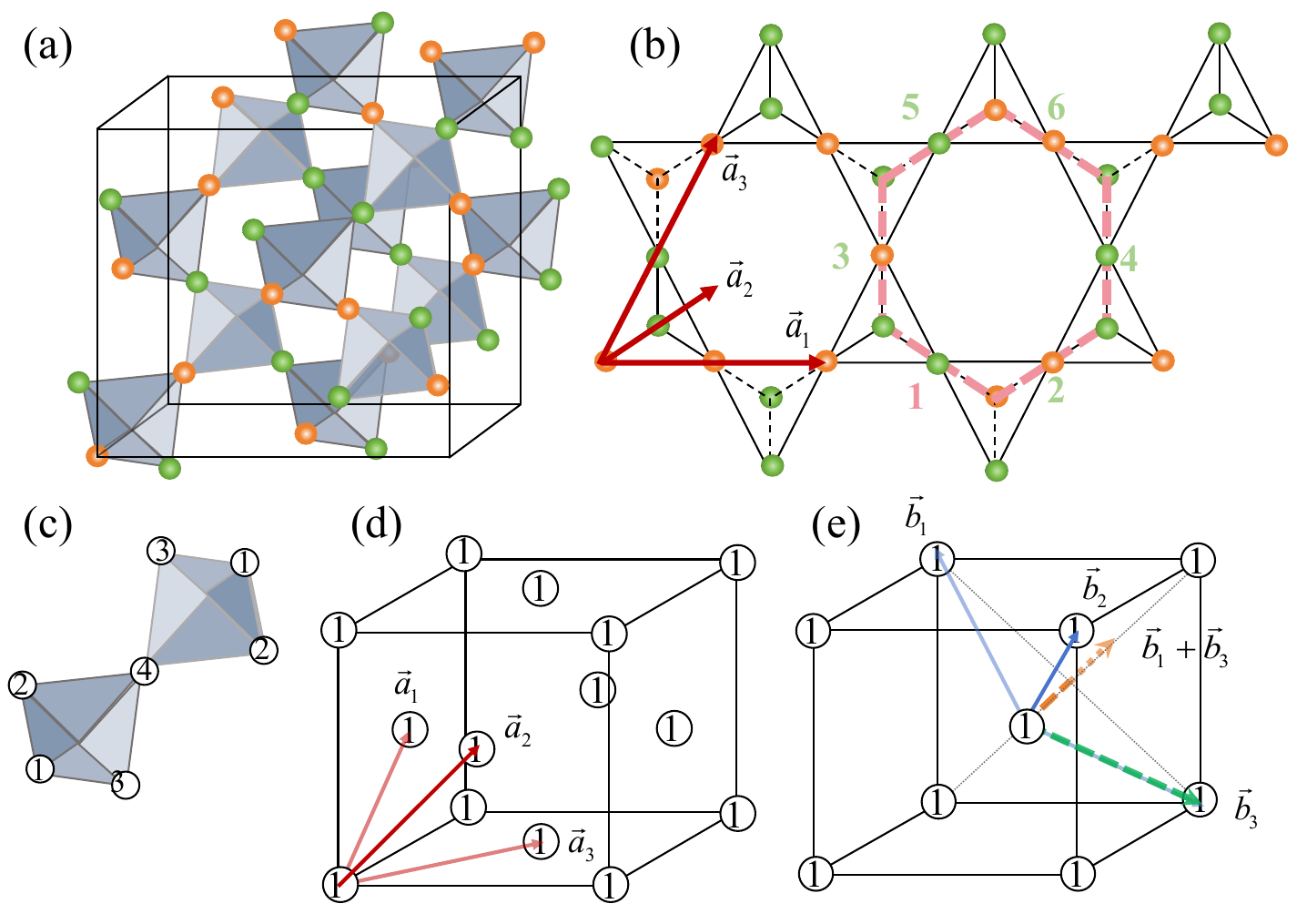}%
\caption{\label{fig:20} Pyrochlore lattice. (a) Each sites in the pyrochlore lattice is shared by an up-pointing tetrahedron and a down-pointing tetrahedron. Different colored dots indicate that the atoms are in ground states or Rydberg states. The configuration satisfies the ice rule. (b) The pink dashed line connects the centers of the nearest-neighbor tetrahedron to form a hexagonal cell. We label the atoms at different positions from 1 to 6 respectively. The atoms in the Rydberg state are mapped to dimers, while the atoms in the ground state are mapped to empty links. Therefore, the Rydberg atomic array on the pyrochlore lattice can be mapped to a QDM on diamond lattice. (c) The pyrochlore lattice is a composite lattice composed of four face-centered cubic (FCC) lattices. Each vertex of a tetrahedron belongs to a different sublattice. The sites in different sublattices are labeled as 1, 2, 3, and 4. (d) For one sublattice, the direct lattice vectors are $\overrightarrow{a_1}, \overrightarrow{a_2}, \overrightarrow{a_3}$. (e) The reciprocal lattice is body-centered cubic (BCC) with vectors $\overrightarrow{b_1}, \overrightarrow{b_2}, \overrightarrow{b_3}$. We measure the dispersion along the $ \overrightarrow{b_1}+\overrightarrow{b_3}$ and $\overrightarrow{b_3}$.} 
\end{figure}

%###################################################
The U(1) Coulomb liquid phase on a 3D pyrochlore lattice has been studied in a long time both in condensed matter and cold-atom systems~\cite{Hermele2004pyrochlore,huang2018dynamics,isakov2005spin,moessner2006quantum,PhysRevX.4.041037,shah2024quantumspinicethreedimensional}. This phase is also referred to as ``U(1) QSI'' in the context of pyrochlore quantum Ising magnets, and can be well understood in terms of the QDM on the dual diamond lattice.
The geometry of the pyrochlore lattice is shown in Fig.~\ref{fig:flux} and Fig. \ref{fig:20}, where each pyrochlore site is shared by one up-pointing tetrahedron and one down-pointing tetrahedron. 
The centers of these tetrahedra form a bipartite diamond lattice. 
For convenience in later discussions, we label the centers of the up-pointing and down-pointing tetrahedra as a and b sublattices of the diamond lattice, respectively. 
Before discussing the low-energy effective model of the Rydberg arrays, let us review the realistic Hamiltonian in Rydberg experiments:  

\begin{alignat}{1}
  \hat H=& \sum_{i=1}^N \left[\frac\Omega 2\left(
  \left|g\right>_i\left< r\right|_i + \left|r\right>_i\left< g\right|_i\right)-\delta\left|r\right>_i\left<r\right|_i\right]\nonumber \\
  &+\sum_{i,j=1}^N \frac {V_{ij}} {2} \left(\left|r\right>_i\left<r\right|_i\otimes\left|r\right>_j\left<r\right|_j\right),
\label{eq:eq1}
\end{alignat}
where the sum on $i$ runs over all $N$ sites of the lattice. The ket $\left|g\right>$ ($\left|r\right>$) represents the ground (Rydberg) state, while $\Omega$ ($\delta$) stands for the Rabi frequency (detuning) of the laser drive. 
The repulsive interaction is of the van der Waals form 
$V_{ij}=C_6/r_{ij}^6 $
where $r_{ij}$ is the distance between the sites $i$ and $j$, and $C_6$ is the van der Waals coefficient.

The above Rydberg model can be mapped to the quantum Ising model with both transverse and longitudinal fields.
We can map the two levels of atoms to effective spin-$1/2$ objects: $\frac{1}{2}\left(\left|g\right>_i\left<r\right|_i+ \left|r\right>_i\left<g\right|_i\right)=\hat{S}_i^x$ and $\left|r\right>_i\left<r\right|_i = \hat{n}_i = \hat{S}_i^z+\frac{1}{2}$. The Hamiltonian takes the same form $H = \hat{H}_0 + \hat{H}_\Omega$ in studies of finite-size systems and thermodynamic limit \cite{Ebadi.2021,Samajdar_2020,cheng2024emergent}, where
%ebadi2021quantum,Ebadi.2021
%Samajdar_2020,samajdar2020complex.
\begin{alignat}{1}
  &\hat{H}_{\Omega} = \sum_i \Omega \hat{S}_i^x , \notag \\ 
  &\hat{H}_0 = - \sum_i h \hat{S}_i^z + \sum_{i,j=1}^N \frac{V_{ij}}{2} \hat{S}_i^z \hat{S}_j^z,
\end{alignat}
with effective local longitudinal field $h=\delta-\frac{1}{2N}\sum_{i\neq j}V_{ij}$. The local field can induce single-spin flips, thereby altering the density of atoms excited to the Rydberg state. In studies of finite-size systems, it is essential to account for the impact of such local effects. To this end, we have examined cases with different filling densities in our subsequent analysis, specifically: only one atom within each tetrahedron is excited to the Rydberg state, and two atoms within each tetrahedron are excited to the Rydberg state.

Notice that the van der Waals interaction $V_{ij}$ term leads to the blockade effect, meaning that Rydberg excitations within the blockade radius $R_b$ experience large pairwise repulsive interactions and hence should be suppressed. The blockade radius is given by $R_b = (C_6 / \hbar \Omega)^{1/6} $, where and $ \hbar$ is the reduced Planck constant. Specifically, when the blockade radius is equal to the distance between nearest-neighbor atoms in the pyrochlore lattice, the lattice strictly satisfies the local constraint (at most one atom per tetrahedron can be excited to the Rydberg state).
The consequence of such Rydberg blockade effect is clearly seen if we only keep the van der Waals interaction at the nearest neighbor $V_1$, where further neighbor interactions are ignored due to the rapid $\sim r^{-6}$ decay of the van der Waals potential. $H_0$ is then expressed as:
\begin{equation}
   H_0=\frac{V_{1}}{2}\sum_\mathsf{r}(\hat{n}_\mathsf{r}-\rho)^2,
\end{equation}   
where the summation is performed on all the dual diamond lattice site $\mathsf{r}$, $\hat{n}_\mathsf{r}=\sum_{i\in \tetrahedron _\mathsf{r}} \ket{r}_i\bra{r}$ is the total Rydberg atom occupancy of the tetrahedron centered at $\mathsf{r}$, and $\rho=2+\frac{h}{2V_{1}}$. 
It is found that the energy of $H_0$ is minimized when each tetrahedron is occupied by $n = \text{floor}(\rho + 1/2)$ Rydberg atoms for $-1/2 <\rho< 9/2 $ 
~\cite{shah2024quantumspinicethreedimensional}. 
This is the origin of local constraint, or the ``ice rule'' $\hat{n}_\mathsf{r}=n$ for each tetrahedron, as imposed dynamically by the van der Waals interactions. 
At finite detuning, more Rydberg excitations are preferred, that seems to contradict with one-Rydberg excitation-per-tetrahedra constraint. 
As a compromise,  it turns out that at low energies, each tetrahedra is occupied by n Rydberg atoms. Another evidence is the nematic phase (two excited Rydberg atoms in nearest neighbor) can be found in Rydberg array \cite
{Samajdar.2021}
%{samajdar2021quantum}. 
Meanwhile, the tetrahedra that do not satisfy the local constraint $\hat{n}_\mathsf{r}=n$ will contribute energy of $\sim V_1$ above the ground state, and are denoted as ``monomer excitations''.

For the limit $ V_{1} \gg \Omega$, $\hat{H}_\Omega$ can be regarded as a perturbation. The low-energy effective Hamiltonian under Schrieffer-Wolff formulation of perturbation theory up to the sixth order can be obtained~\cite{shah2024quantumspinicethreedimensional}. 
We will describe the perturbative process using the language of hard-core boson, which will hereafter be referred to simply as 'boson'. 
To simplify the discussion, we derive here the effective Hamiltonian for the case of n=2 (i.e., two dimers per site). Since the creation or annihilation of a boson on a single lattice site would violate the ice rule, it would lead to the appearance of either three dimers per site or one dimer per site.
So the first-order contribution clearly vanishes. A similar situation arises at all odd orders, so we only consider corrections at even orders. 
At the second order, bosons must be created and annihilated at a single site to return to the ice manifold. This process does not alter the overall configuration of the system.
This can occur at any point, contributing a constant term to the energy. 
At the fourth order, bosons hop along a link to a neighboring site and then return along the same link, contributing a constant energy. Such processes are present for all links. This process does not change the positions of the bosons and, from the perspective of the dual lattice, the local constraint of two dimers per site remains satisfied.
At the sixth order, one process involves a single boson hopping around the triangular face of each tetrahedron. 
Another process, as shown in Fig. \ref{fig:20}(b), features atoms alternating between the ground and excited states along a hexagonal cell. This alternating configuration can be mapped, via a ring-exchange process, onto a complementary and flippable configuration in the dimer model. This process does not violate the local constraint (two dimers per site). 
The former contributes a constant energy.  Ignoring all these constant terms, the effective Hamiltonian is a six-site ring exchange term
\begin{equation}
   H_\textrm{eff}^{(6)}=-t\sum \left( S_1^+ S_2^- S_3^+ S_4^- S_5^+ S_6^-  +h.c. \right) ,
\label{eq:eq4}
\end{equation}
where the sum is over all the hexagonal cells in the pyrochlore lattices. $t$ can be obtained by summing over virtual processes and $S_i^\pm$ represent that boson (spin) creation or annihilation at a single site $i$. Although this effective Hamiltonian is derived in the specific case of n=2, it is worth noting that for n=1 and n=3, the effective Hamiltonian we obtain is all described by the same universal form, namely Eq.~\eqref{eq:eq4}.

The Schrieffer-Wolff perturbation theory in the context of quantum spin ice has also described in the paper \cite{shah2024quantumspinicethreedimensional}.
%\cite{Hermele2004pyrochlore}. 
This perturbation theory holds as long as the low-energy manifold hosts fixed number of Rydberg atoms within each tetrahedra (ice rule). For finite systems, as long as
the tetrahedra structures are complete at the boundaries, the Rydberg blockade mechanism guarantees the
low-energy configurations to obey such ice rule. As a result, the Schrieffer-Wolff formalism still applies
and our results are applicable to finite-size spin systems given that tetrahedra structures are complete
along the boundaries.
 
For mapping to a QDM on the dual lattice, we connect the centers of all nearest-neighbor tetrahedra as shown in Fig. \ref{fig:flux} (b) and \ref{fig:20} (b), with each atom exactly at the center of each link. Atoms in the Rydberg state (spin up) can be mapped to dimers in the dual (diamond) lattice, while atoms in the ground state (spin down) can be mapped to empty links. The high-order ring exchange dynamics becomes the effective kinetic term. Thus the QDM on diamond lattice takes the role of low-energy effective description of this pyrochlore Rydberg array~\cite{shah2024quantumspinicethreedimensional}.
The ground state of quantum dimer models can be exactly solved by adding a potential energy term $V$ and adjusting it to a particular value $V=t$, referred to as the RK point~\cite{moessner2011quantum}. At the microscopic level, this $V$ term corresponds
to high-order density interactions between Rydberg atoms and is not realistic in Rydberg experiments.
However, in the context of pyrochlore quantum spin ice, it is reported in \cite{PhysRevLett.127.117205} that the third-nearest-neighbor Ising interaction $J_3$ plays a similar role to $V$, that can tune the system to the field-theoretical RK point with vanishing photon velocity. In Rydberg platform, this $J_3$ term also presents as the third-nearest-neighbor van der Waals interactions between Rydberg atoms.

The Hamiltonian of QDM with the potential energy term on the dual diamond lattice is
\begin{equation}
\includegraphics[width=0.3\textwidth]{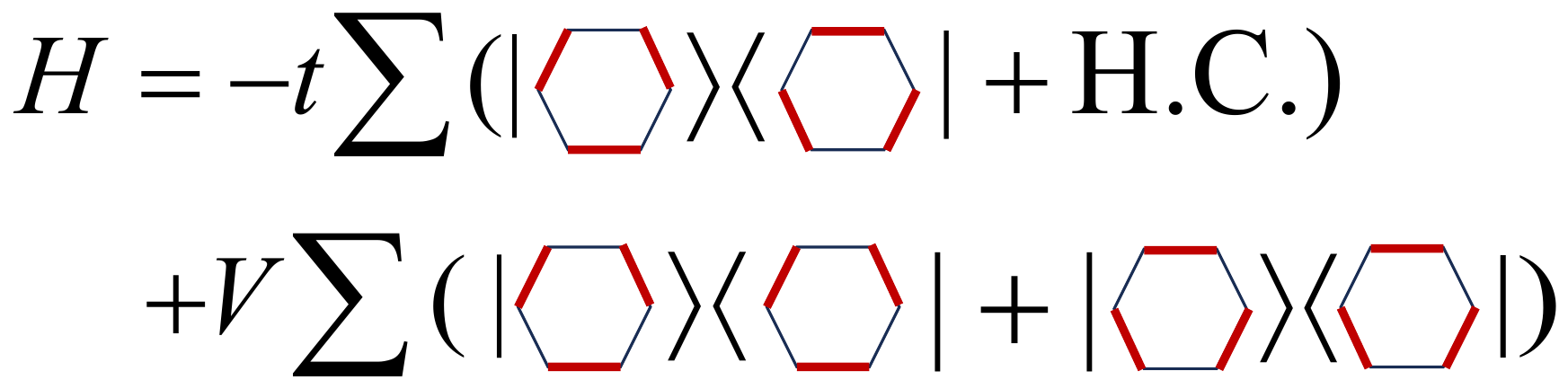},
\end{equation}
where the sum is performed over all hexagonal plaquettes. The first term is the ring exchange term and the second term is potential energy term. 

When $t/V=1$, \textit{i.e.}, at the famous RK point~\cite{moessner2011quantum}, the ground state
is a resonating valence bond (RVB) state described by the coherent equal-weight superposition of all the dimer covering configurations, \textit{i.e.}, $|G\rangle=\sum_C |C\rangle$. Moreover, the RK point exhibits topological degeneracy, where the equal-weight superpositions within different topological sectors are ground states with identical energies~\cite{ZY2020improved}.

In the phase diagram of QDMs on 3D bipartite lattices, the RK point is located as a part of the U(1) Coulomb phase. 
In the spin language, the local constraint of dimers corresponds to the ice rule that the total spin in each tetrahedron of the pyrochlore lattice is same. 
In the dimer language, it means that the number of dimers attached to each site of the dual diamond lattice is fixed.
Further, this local constraint is an essential ingredient that allows a mapping to the U(1) gauge theory: 
Spin-up (dimer) and spin-down (empty link) can be mapped to the $\pm\frac{1}{2}$ magnetic flux
pointing from a to b diamond sublattices,
as displayed in Fig. \ref{fig:flux} \cite{PhysRevX.4.041037, PhysRevB.90.245143}. 
This leads to the condition $\left(\mathrm{div} B \right)_\mathsf{r} = 0$ for the case of two-dimers-per-site configurations, and $\left(\mathrm{div} B \right)_\mathsf{r} = \eta_\mathsf{r}$ for the one-dimer-per-site configurations. 
Here $\mathsf{r}$ denotes the site of the dual diamond lattice, and $\eta_\mathsf{r}=\pm 1$ for a and b sublattices, respectively. 
Such divergence condition is analogous to Gauss's law of electromagnetism, and the background charges can be absorbed by redefining the strength of the magnetic field.  

The pyrochlore lattice is a composite, consisting of four sets of face-centered cubic lattices. As shown in Fig.~\ref{fig:20} (c), we label the points on the tetrahedron belonging to different sublattices as 1, 2, 3, and 4. In Fig.~\ref{fig:20} (d), we display one of the sublattices and define its direct lattice vectors as $\overrightarrow{a_1}, \overrightarrow{a_2},  \overrightarrow{a_3} $, with the corresponding reciprocal lattice vectors as $\overrightarrow{b_1}, \overrightarrow{b_2}, \overrightarrow{b_3}$ (see Fig.~\ref{fig:20} (e)).

Because the ground state wave-function at the RK point is an equal-weight superposition of all possible configurations, i.e., $\ket{G}=\sum_C \ket{C}$, the expectation value of any physical observable $O$ can be sampled as
\begin{equation}
   \langle O \rangle=\frac{\langle G|O|G\rangle}{\langle G|G\rangle} = \frac{\sum_{C,C'}\langle C|O|C' \rangle}{\sum_{C,C'} \langle C|C' \rangle}.
\label{6}
\end{equation}
Considering the orthogonality of the dimer basis, the equation becomes
\begin{equation}
   \frac{\langle G|O|G\rangle}{\langle G|G\rangle} = \frac{\sum_{C}\langle C|O|C' \rangle}{\sum_{C} \langle C|C \rangle}.
\label{7}
\end{equation}
In this case, we only sample $C$, because once $C$ is fixed, there is only one $C'$ such that the matrix element $\langle C'| O | C \rangle \neq 0$.
Therefore, the expectation value of observables could be gained via sampling different configurations $C$ with equal weight. That is, randomly sample the configurations $C$ and average the values $\langle C|O|C' \rangle$ to gain $\langle O \rangle$.

Through the above way, the static properties can be calculated and explored. For the dynamical excitations above the RK ground state, 
the single mode approximation (SMA) is a useful tool for analyzing the low-lying dispersion of the model~\cite{feynman1954atomic,griffin2009bose,toennies2004superfluid,haegeman2013elementary,yi2002single,bruschi2010unruh,Lauchli2008SAC,yan2022height,liu2024accessingexcitationspectrummanybod},
which transforms the problems of solving dynamical properties into evaluations of equal-time correlations, as shown in Eq. \eqref{7}.
%as can be achieved in Eq. \eqref{eq:eq5}. 
For example, for the dynamical dimer density correlation, we assume that the excited-states are described by the single-mode ansatz: $|\textbf{q},\alpha\rangle=D_\alpha(\textbf{q})|G\rangle$, then then the upper bound of the excited state energy $\omega_{SMA}$ can be obtained as:
\begin{equation}
    \omega_{SMA}=\frac{1}{2} \frac{\langle G|[D_\alpha (-\textbf{q}) ,[H, D_\alpha (\textbf{q})]]|G \rangle}{\langle G| D_\alpha (-\textbf{q}) D_\alpha (\textbf{q}) |G \rangle},
\label{eq:eq2}
\end{equation}
where $D_\alpha(\mathbf{q})=\frac{1}{\sqrt{N}}\sum_{\mathbf{r}} e^{-i\mathbf{q}\cdot\mathbf{r}}D_\alpha(\mathbf{r})$ is the Fourier transformed dimer density operator in momentum space, and $D_\alpha(\mathbf{r})$ is the dimer density operator in the real space. Here, each dimer is labeled by its unit cell position with the sublattice index $\alpha$ and position \textbf{r}, respectively. We only consider the dispersion in the same sublattice here.

\begin{figure} [h]%tbp]
\includegraphics[width=0.5\textwidth]{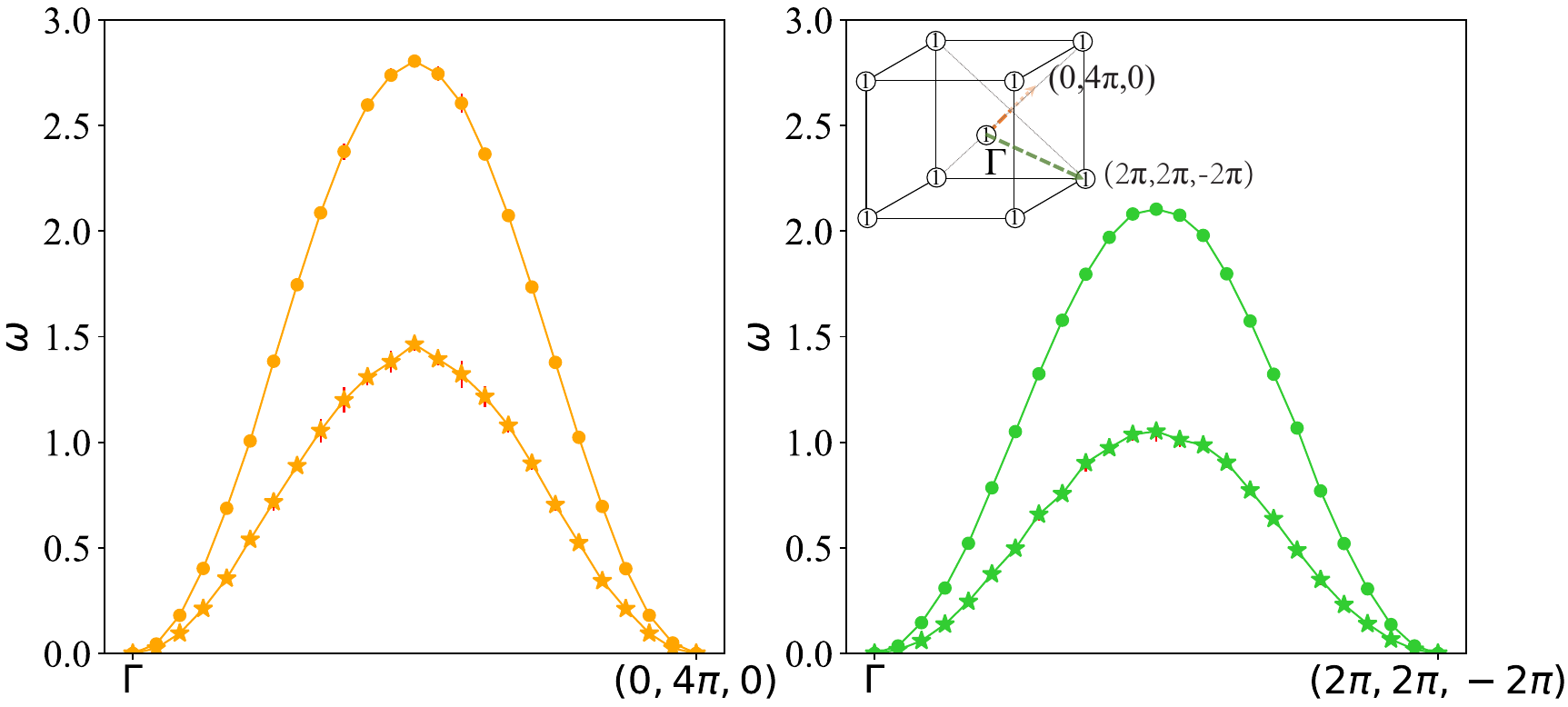}%[scale=0.5]% Here is how to import EPS art
\caption{\label{fig:21}{Dispersion of dimer density correlation for the original RK-QDM obtained by SMA for the diamond lattice ($24\times 24\times 24$) with strict one dimer per site (star) and two dimers per site (circle) along the Brillouin zone path $\Gamma (0, 0, 0) \rightarrow (0,4\pi,0) $  and $\Gamma \rightarrow (2\pi,2\pi,-2\pi)$. }}
\end{figure}

Based on SMA, we consider a $24\times 24\times 24$  diamond lattice with perfect one dimer per site and two dimers per site, and evaluate the dimer dispersion along the direction of $\overrightarrow{b_1}+\overrightarrow{b_3}$ and $\overrightarrow{b_3}$ in momentum space based on Eq.~\eqref{eq:eq2}. The computational approach is to sample the dimer configurations with equal-weight through a random walk. All the updates are local loops which cannot change the winding sector of the dimer configurations. This process is repeated $10000$ times, and the results are statistically averaged to approximate the dispersion of the pyrochlore lattice, $\omega_{SMA}$. We calculate the dispersion of each of the four sublattices separately, and then average them to obtain the dispersion of the entire system. The result is plotted in Fig.~\ref{fig:21}. 
The quantum excitation at the RK point exhibits only one gapless points located at $\Gamma (0, 0, 0)$ (also equivalent to $(0, 4\pi, 0)$ and $(2\pi, 2\pi, -2\pi)$). This only gapless mode corresponds to the photon excitations as predicted by theory. Since the RK point is described by the Lifshitz's theory with dynamical exponent $z=2$, its dispersion is quadratic~\cite{fradkin2013field,moessner2011quantum}.
\begin{comment}

\end{comment}

If the configurations do not perfectly carry fixed integer numbers of dimers per site, then the doping case can be discussed.
For example, when we add dimers into the system of one dimer per site and set the soft constraint with one or two dimers per site only, the site with two dimers can be treated as a doped point. The density of doping is defined as the percentage of the number of doped sites that make up the ground state relative to the total number of lattice sites. In the following, we will tune the local filling from one-dimer-per-site to two, even to three. 
Now we have to define the monomers under doping. If a site violates the local dimer constraint of the reference, it is a monomer. In the paper, when doping density is less than $50\%$, the rule for choosing the reference filling is: one dimer per site if the local dimer filling is one or two; two dimers per site if the local filling is two or three; three dimers per site if the local filling is three or four. When the doping density is greater than $50\%$, we choose the upper local filling as the reference. For example, for one-dimer-per-site configuration with doping density $>50\%$, we choose the reference filling as two dimers per site, the site with one dimer becomes a monomer.

In the gauge field language, monomers are regarded as magnetic charge fluctuations with respect to the background reference configurations. 
If we incorporate the quantum dynamics of monomers as their hopping, we can construct an RK-like effective Hamiltonian based on the original RK-QDM, supplemented by the dimer hopping term and the related diagonal term:
\begin{equation}
\includegraphics[width=0.35\textwidth]{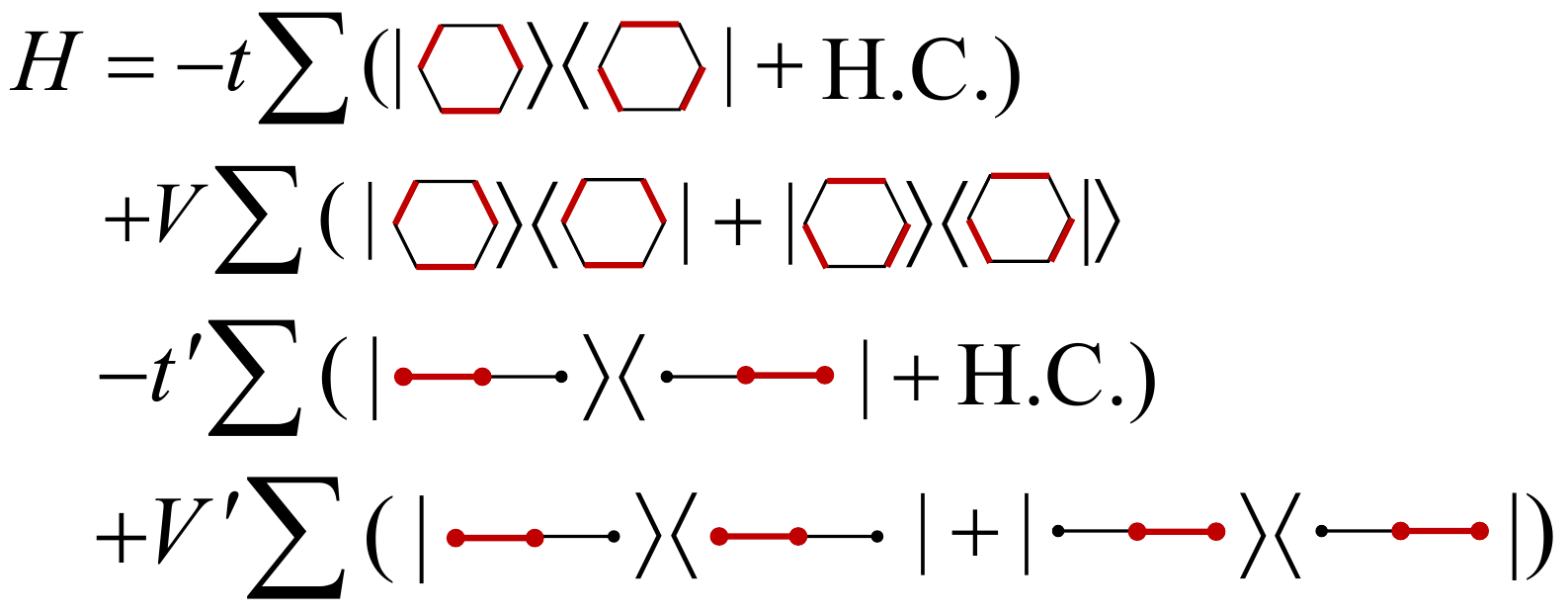},
\label{eq:doped_QDM}
\end{equation}
where the $t'$ and $V'$ terms run over all the nearest-neighbor links of the lattice. The third term is the kinetic energy term for exchanging dimers between adjacent links with the coupling strength $-t^{\prime}$, and the fourth term is the corresponding potential energy term with the coupling strength $V^{\prime}$. 
This system exhibits a global U(1) symmetry associated with conservation of total dimers:
\begin{equation}
    M=\sum_{\textbf{r}\alpha} D_\alpha (\textbf{r}),
\end{equation}
which divides the Hilbert space into sectors characterized by different total dimer occupation $M$. 
When these parameters are tuned to a general RK point ($t=V$ and $t^{\prime}=V^{\prime}$), the ground state wavefunction can be expressed as an equal-weight superposition of connected configurations. The connected configurations mean they can be connected by local terms of the Hamiltonian. Notice that the ground states are same if $t=V$ and $t^{\prime}=V^{\prime}$, but the excitation is dependent on the ratio of $V/V'$. In the following, we fix $t=t'=V=V'=1$ to explore the ground state and excitation properties.

\begin{figure} [h]%tbp]
\includegraphics[width=0.5\textwidth]{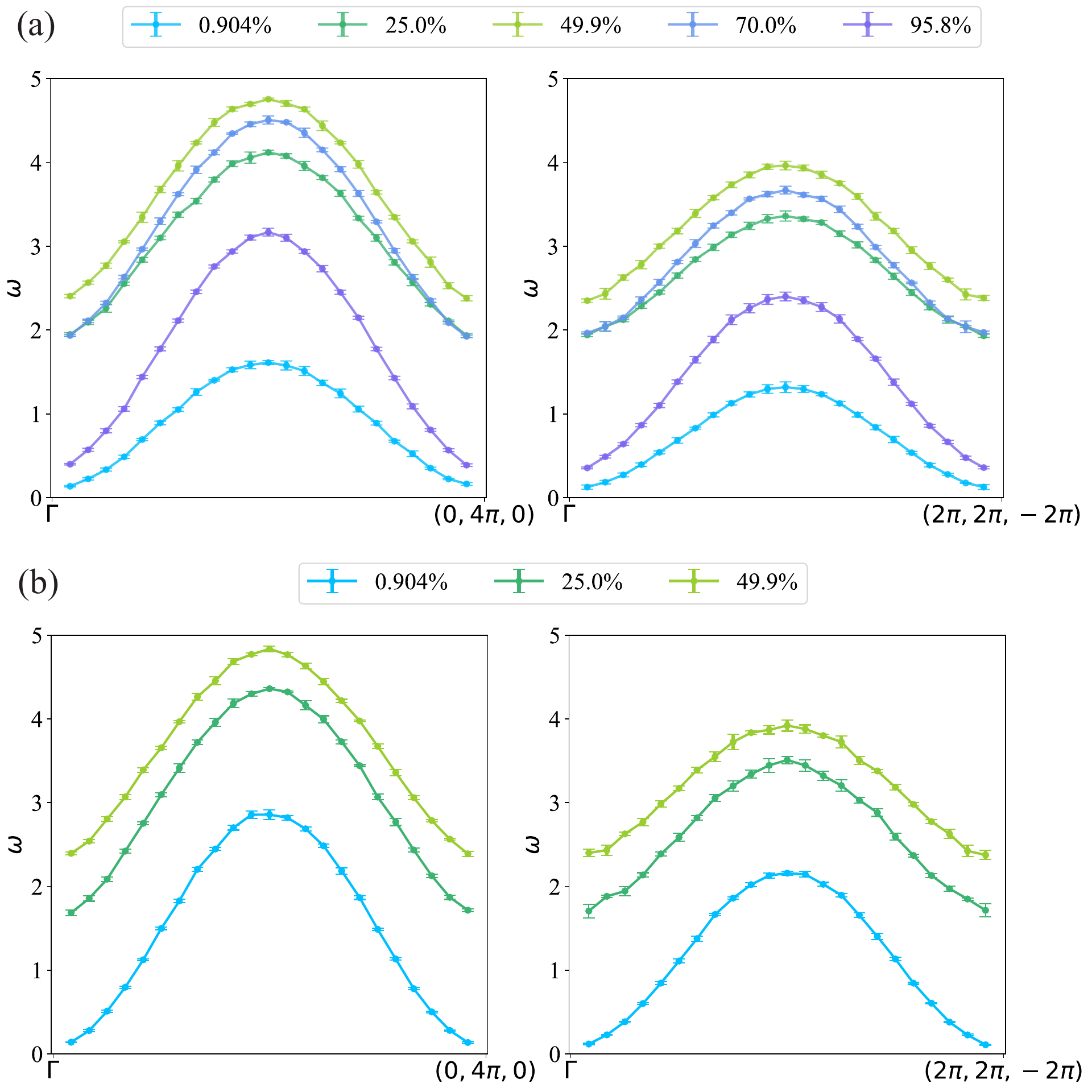}%[scale=0.5] 
\caption{\label{fig:dopsma} The dispersion of dimer density operator of the doped QDM at the RK point  $t=t'=V=V'=1$. We take the Brillouin zone path $\Gamma~(0,0,0) \rightarrow (0,4\pi,0) $ and $\Gamma \rightarrow (2\pi,2\pi,-2\pi)$ on a diamond lattice with system size  $24\times 24\times 24$. 
Different colors correspond to (a) one dimer per site with varying doping densities ($0.904\%$, $25.0\%$, $49.9\%$, $70.0\%$ and $95.8\%$) and (b) two dimers per site with varying doping densities ($0.904\%$, $25.0\%$, and $49.9\%$) .
}
\end{figure}

We further investigate the dynamical dimer-dimer correlation of the doped QDM Eq.\eqref{eq:doped_QDM} under the one dimer per site with varying doping densities ($0.904\%, 25.0\%, 49.9\%, 70.0\%$ and $95.8\%$) and two dimers per site with varying doping densities ($0.904\%, 25.0\%$ and $49.9\%$), see Fig.~\ref{fig:dopsma}.
The results indicate that doping of monomers gaps out the photon excitations at the $\Gamma$ point by Higgsing the emergent U(1) gauge field. Consequently,  the Coulomb liquid phase vanishes and the system should become trivially confined upon doping. 

\begin{figure} [htbp]
\includegraphics[width=0.4\textwidth]{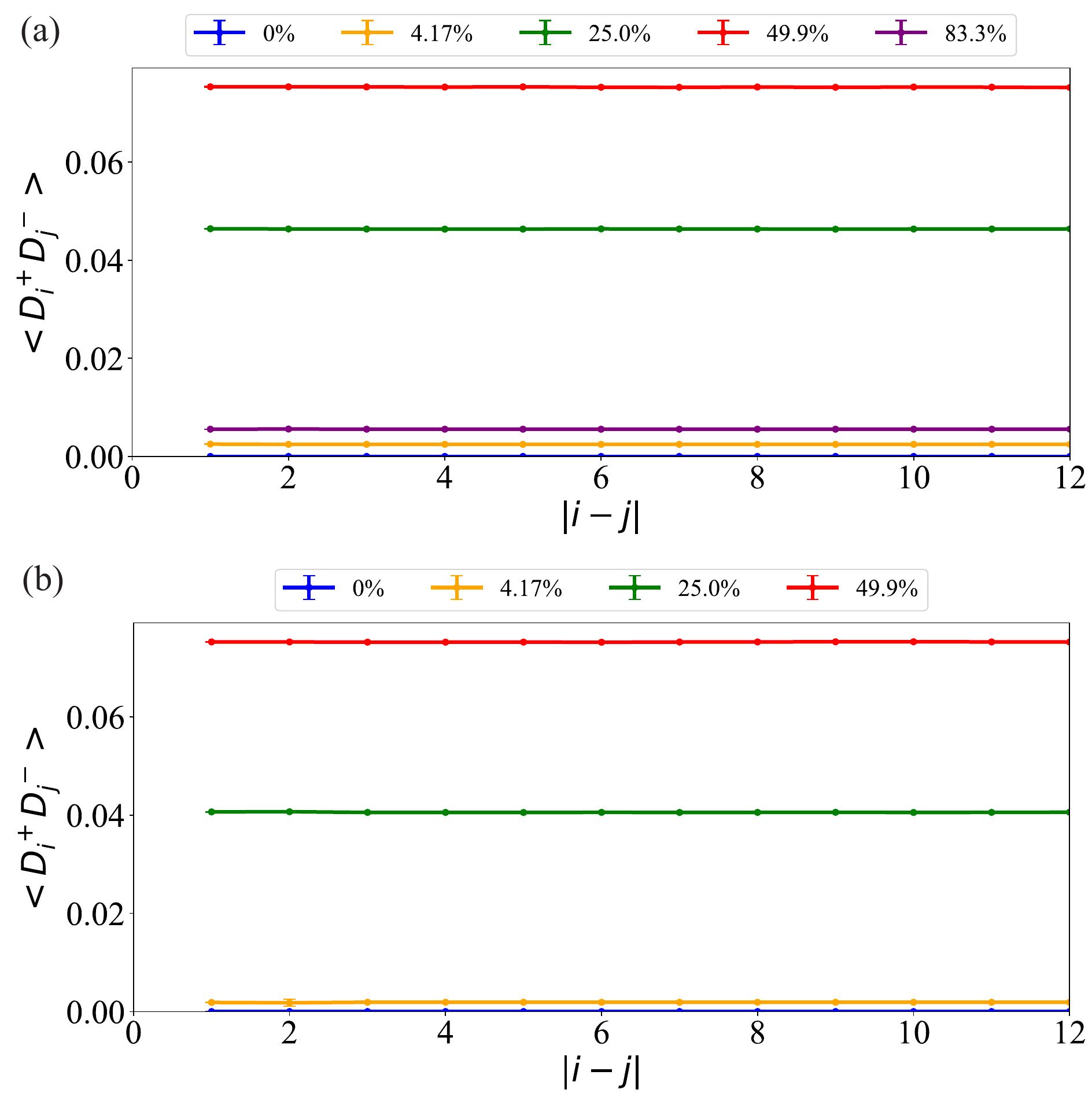}%
\caption{\label{fig:22} Off-diagonal dimer correlation of the doped QDM at the RK point $t=t'=V=V'=1$. The circles represent the off-diagonal correlations along the $\overrightarrow{a_1}$ direction as a function of the distance $ |i-j|$, Different colors correspond to (a) one dimer per site with the varying doping density ($0\%, 4.17\%, 25.0\%, 49.9\%$ and $83.3\%$) and (b) two dimers per site with the varying doping density ($0\%, 4.17\%, 25.0\%, $ and $49.9\%$)
, where $ 0\%$ corresponds to applying a strong constraint to the system that prohibits the presence of monomers. Here, the distance between nearest-neighbor links in the sublattice is set to 1. The values of the off-diagonal dimer correlation are significantly larger than their corresponding errors, the error bars are obscured by the circular data markers.}
\end{figure}

To further explore the physical properties of the QDM upon doping, we evaluate the off-diagonal dimer correlations of the doped QDM by taking $O=D_i ^+D_j ^ -$ in Eq.~\eqref{7}. 
This dynamical correlation function measures excitations involving creation or annihilation of dimers in the RK ground state. 
Here, we set the distance between nearest-neighbor links in the same sublattice to 1. 

In Fig.~\ref{fig:22}, we present the off-diagonal correlations along the $\overrightarrow{a_1}$ direction in one sublattice, under the background of one-dimer-per site and two-dimers-per-site, and compare the impact of different doping densities ($4.17\%, 25.0\%, 49.9\%, 83.3\%$ and $0\%$) on the off-diagonal correlation. 
We did not include the point at $|i-j| = 0$ ($\langle D_i^+D_j^-\rangle  = 1$) in the figure. This omission is intentional, as the steep drop from $i-j = 0$ to $i-j = 1$ would cause all the curves for $i-j \geq 1$ to collapse onto each other, which would obscure the differences between curves for different doping densities. 

In addition, when $i - j \geq 1$, the value of the off-diagonal dimer correlation $\langle D_i^+ D_j^- \rangle$ is independent of the distance $|i - j|$. The underlying reason for this phenomenon is that the distribution of monomers within the same sublattice is statistically uncorrelated. %independent.  
According to Eqs. \eqref{6} and \eqref{7}, only dimer configurations $C$ with site $i$ not attached to any monomers and site $j$ attached to two monomers has non-vanishing contribution to $\langle D_i^+ D_j^- \rangle$. 
Therefore, $\langle D_i^+ D_j^- \rangle$ essentially measures the statistical weight of such configurations. 
Given that the wave function at the generalized RK point is the equal-weight superposition of all allowed dimer configurations, 
As a result, monomers are spatially independently distributed, leading to a correlation function that is independent of $|i - j|$, and hence a flat curve.

The results show that the off-diagonal correlation vanishes in the undoped case, 
as the dimer hopping term $D_i ^+D_j ^ -$  breaks up the one-dimer-per-site local constraint.
In contrast, while such local constraints become soft upon doping, 
the system immediately develops non-vanishing dimer off-diagonal long-range correlations.
This presence of off-diagonal long-range order can be understood as condensation of monomers that disrupts the local constraint and breaks the U(1) gauge field, hence the Coulomb liquid breaks down.

\begin{figure} [htbp]
\includegraphics[width=0.45\textwidth]{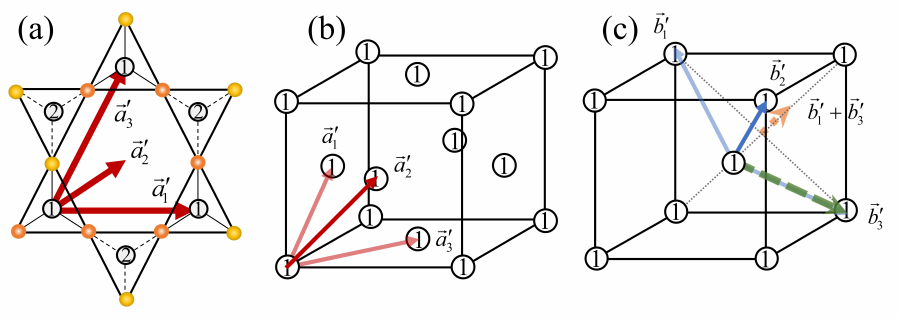}%
\caption{\label{fig:1-4} The vector directions associated with the dual (diamond) lattice.  (a) The Rydberg array located on the sites of the pyrochlore lattice can be mapped to a dimer on a link of the dual (diamond) lattice which is composed of two face-centered cubic lattices. The sites of the dual (diamond) lattice are the body centers of each upward and downward tetrahedron, and they belong to different sublattices. We label the different sublattices as 1 and 2. (b) For one sublattice, the direct lattice vectors are $\overrightarrow{a}_1^{'}, \overrightarrow{a}_2^{'}, \overrightarrow{a}_3^{'}$. (c) The reciprocal lattice is body-centered cubic (BCC) with vectors are $\overrightarrow{b}_1^{'}, \overrightarrow{b}_2^{'}, \overrightarrow{b}_3^{'}$.
 }
\end{figure}

\begin{figure} [htbp]
\includegraphics[width=0.45\textwidth]{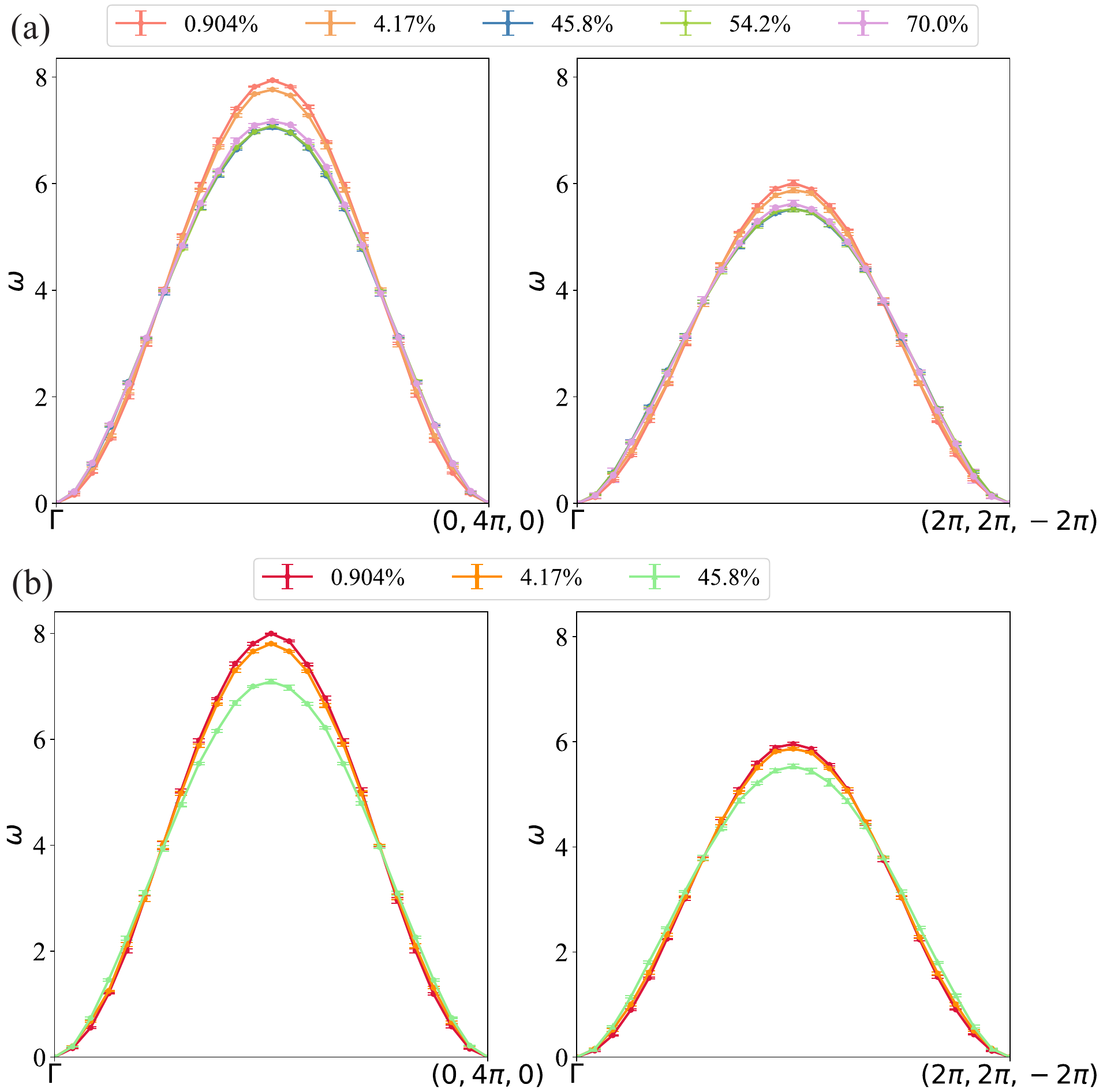}%
\caption{\label{fig:monsma} Dispersion of monomer density correlation at the RK point ($t=t'=V=V'=1$) along the Brillouin zone path $\Gamma \rightarrow (0,4\pi,0) $ and $\Gamma \rightarrow (2\pi,2\pi,-2\pi)$ on a $24\times 24\times 24$ diamond lattice. Different colors correspond to (a) one dimer per site with doping densities ($0.904\%, 4.17\%, 45.8\%, 54.2\%$ and $70.0\%$) and (b) two dimers per site with doping densities ($0.904\%, 4.17\%$ and $45.8\%$).}
\end{figure}

To investigate the quantum dynamics of monomers, we also evaluate the dispersion of monomer density operator based on the SMA. Similarly, we define an excited state $|\textbf{q},\alpha\rangle'=m_\alpha(\textbf{q})|G\rangle$, from which the upper bound of the excitation energy $\omega_{SMA}$ can be obtained as: 
\begin{equation}
\omega_{SMA}^\prime=\frac{1}{2} \frac{\langle G|[m_\alpha (-\textbf{q}) ,[H, m_\alpha (\textbf{q})]]|G \rangle}{\langle G| m_\alpha (-\textbf{q}) m_\alpha (\textbf{q}) |G \rangle},
\label{eq:eqm}
\end{equation}
where $m(\mathbf{q})=\frac{1}{\sqrt{N}}\sum_{\mathsf{r}} e^{-i\mathbf{q}\cdot\mathsf{r}}m_\alpha(\mathsf{r})$ represents the density of monomer in momentum space, and  $\frac{1}{\sqrt{N}}$ is the normalization constant. $m_\alpha(\mathsf{r})$  is the density of monomer in real space, where the label $\alpha $ refers to a specific sublattice of the composite lattice (Fig.~\ref{fig:1-4}). $m_\alpha(\mathsf{r})=1$ means that the site at position $\mathsf r$ in a given configuration $ C $ violates the background constraint, while $m_\alpha(\mathsf{r})=0$ indicates that the site at position $\mathsf r$ in configuration $c$ follows the background constraint. Here, we also consider a $ 24 \times 24 \times 24 $ dual diamond lattice with local constraint based on the Eq.~\eqref{eq:eqm}.

In  Fig.~\ref{fig:monsma}, we present the dispersions of monomers based on one-dimer-per-site configurations with varying doping density ($0.904\%, 4.17\%, 45.8\%, 54.2\%$ and $70.0\%$) and two-dimers-per-site configurations with varying doping density ($0.904\%, 4.17\%$ and $45.8\%$). The results reveal gapless states at the $\Gamma$ point. This gapless mode is identified as the Goldstone mode associated with the monomer condensates that also order at the $\Gamma$ point. This phase is quite analogous to superfluids as it also exhibits a gapless Goldstone mode associated with U(1) symmetry breaking. However, it is still different from superfluids, as the Goldstone mode here exhibits a quadratic dispersion with vanishing superfluid stiffness, and is unstable against viscosity.

%###################################################
\section{Cubic lattice}
\label{section:Cubic lattice}
%###################################################
Then we consider another three-dimensional general RK model where each atom is positioned on the link of cubic lattice. 
This setting is the generalization of the Rydberg implementation of the two-dimensional quantum square ice model into the three-dimensional cubic lattice~\cite{PhysRevX.4.041037}. It's worth noting that, unlike the diamond lattice general RK model, this model is not easy to realize through Rydberg arrays, but its physics is similar to the diamond one. Due to the simpler lattice structure, more physics can be explored here. First, we go back to the Hamiltonian of pure RK model in this setup, which is defined as follows:
\begin{equation}
\includegraphics[width=0.3\textwidth]{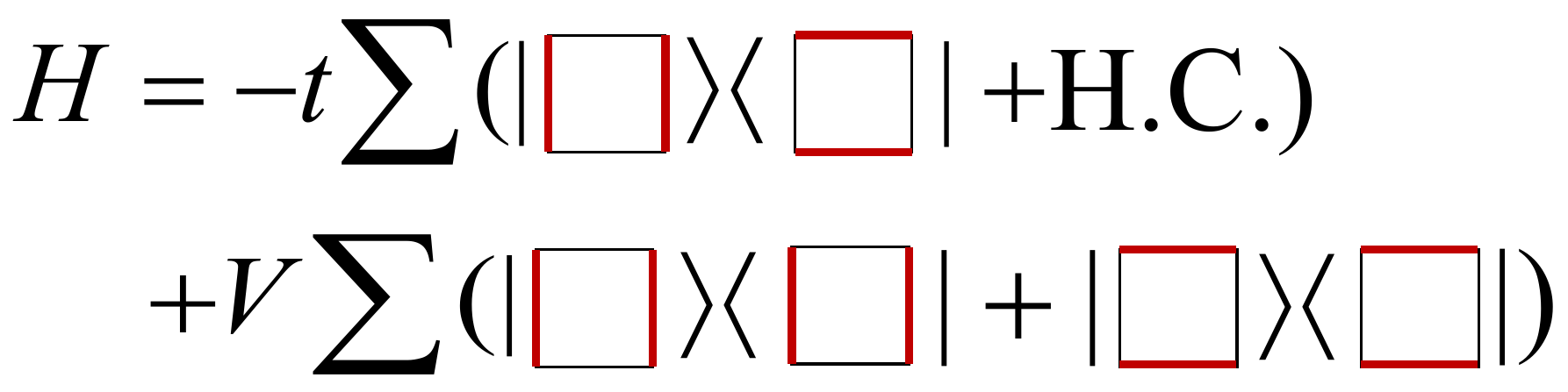},
\label{eq:QDM_cubic}
\end{equation}
where the summation is over all plaquettes within the cubic lattice. The first term is the kinetic term which represents the ring-exchange process on flippable plaquettes, 
while the second term counts the total number of flippable plaquettes. 
At the RK point $V=t$, the ground state is exactly solvable as described by equal-weight superposition of all dimer configurations within each topological sector. 

We evaluate the dispersion of cubic lattice QDM at the RK point $V=t=1$ based on SMA, with the computational approach similar to that described in Sec. \ref{section:Pyrochlore lattice}. 
The system size of the cubic lattice in our simulation is $32\times 32\times 32 $. 
Considering the cubic symmetry of the lattice structure, without loss of generality, we only consider the correlations of dimer density along the $\overrightarrow{x}$ direction $D_x$.

\begin{figure} [htbp]
\includegraphics[width=0.45\textwidth]{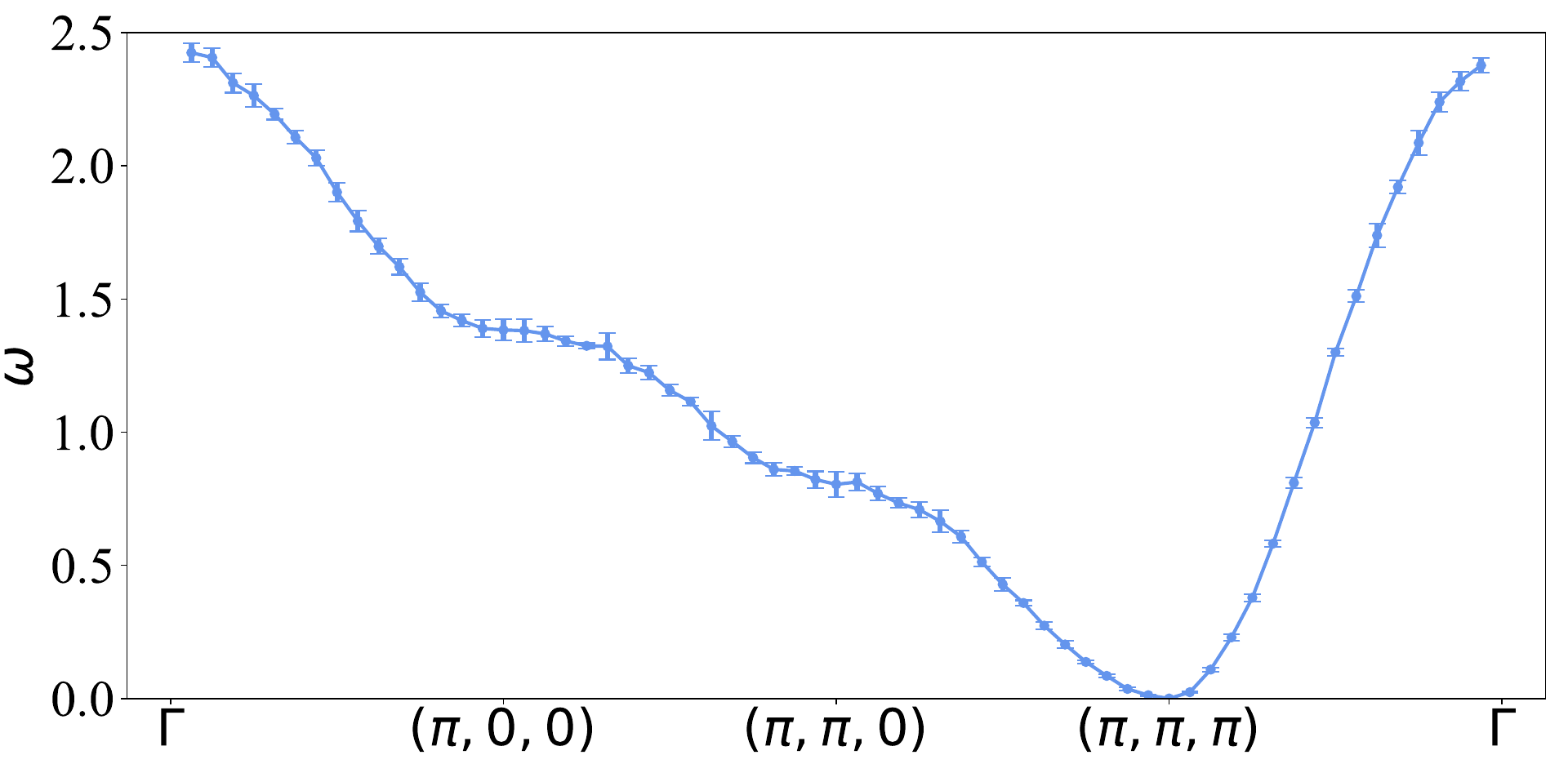}%
\caption{\label{fig:1} Dispersion of dimer of the cubic lattice 
at the RK point $V=t=1$. Under the local constraint of one dimer per site, the dispersion of the QDM on a cubic lattice based on the SMA along the path $\Gamma \rightarrow (\pi,0,0) \rightarrow (\pi,\pi,0) \rightarrow  (\pi,\pi,\pi)\rightarrow  \Gamma$.}
\end{figure}

We start with the case where one-dimer-per-site local constraint is strictly satisfied with no monomers present. We calculate the dispersion relation associated with the dimer density correlation, 
and the results are shown in Fig.~\ref{fig:1}. 
In the dispersion relation, we find only one gapless mode at the $(\pi,\pi,\pi)$ point with quadratic dispersion~\cite{Lauchli2008SAC,moessner2003three,huse2003coulomb}. 
This result is consistent with the theoretical prediction that such RK point lies as part of the U(1) Coulomb liquid phase and hosts a gapless photon mode at $(\pi,\pi,\pi)$, with the speed of light vanishes right at the RK point.

Then we dope the system with monomers. To incorporate monomer quantum dynamics, we introduce some monomer hopping terms above the original RK-QDM, where the Hamiltonian becomes:

\begin{equation}
\includegraphics[width=0.4\textwidth]{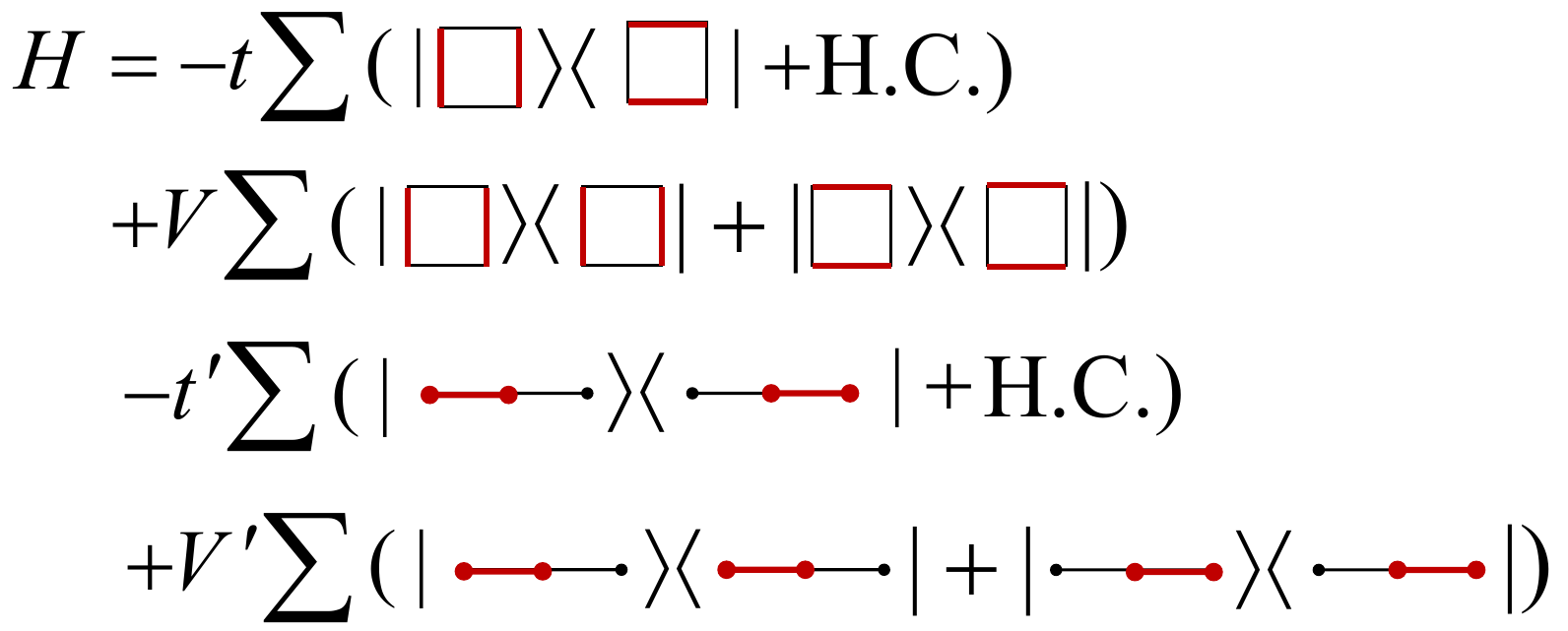},
\label{eq:doped_QDM_cubic}
\end{equation}
where the summation of the third and the fourth terms are performed throughout all links, the third term 
is the dimer hopping term that incorporates hopping of monomers between different sites, and the fourth term 
counts the total number of hoppable dimers. 

This doped QDM Eq. \eqref{eq:doped_QDM_cubic} also exhibits a global U(1) symmetry associated with conservation of total dimers:
\begin{equation}
    M=\sum_{\textbf{r},\alpha} D_\alpha (\textbf{r}).
\end{equation}
This symmetry divides the overall Hilbert space into different sectors characterized by their total dimer occupation $M$. 
Within each sector, the ground state is also exactly soluble at the generalized RK point $t=V$ and $t' =V'$, with energies being zero, and the wave functions described by equal-weight superposition of all configurations within the sector. 
Without loss of generality, in the following we only present results at a specific point $t=V=t' =V'=1$.
As discussed previously, monomers are defined with respect to the closest background configurations that satisfy local constraints.
For example, if the doping density of one-dimer-per-site configurations exceeds 50\%, we adjust the background to configurations with two dimers per site to define the monomers.

\begin{figure} [ht]
\includegraphics[width=0.45\textwidth]{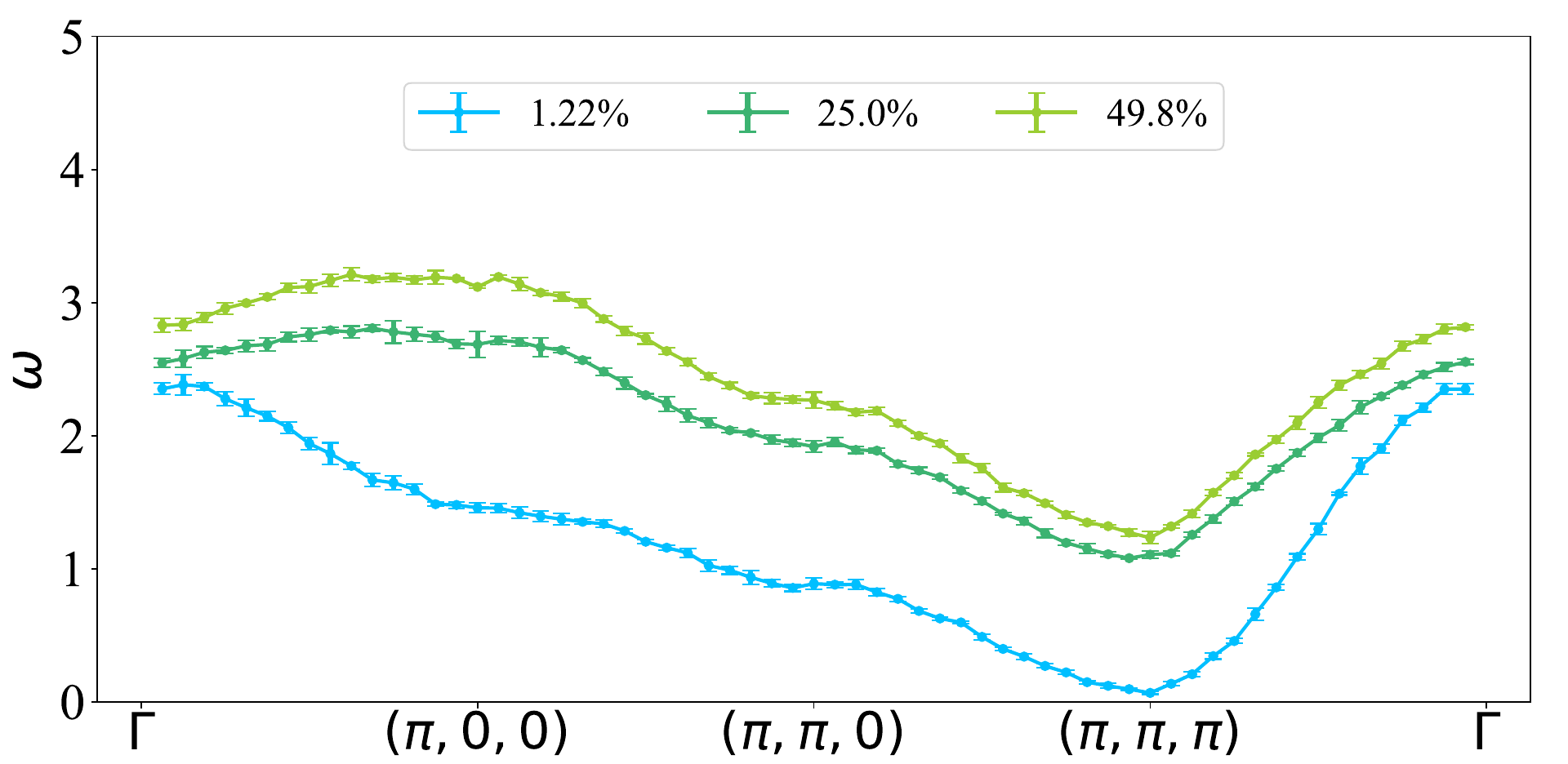}%
\caption{\label{fig:dimer-sma}  Dynamical dimer density dispersion on a cubic lattice with doping
along the Brillouin zone path $\Gamma \rightarrow (\pi,0,0) \rightarrow (\pi,\pi,0) \rightarrow  (\pi,\pi,\pi)\rightarrow  \Gamma$. The different colored circles correspond to the doping densities $1.22\%, 25.0\%$ and $49.8\%$ in one dimer per site. }
\end{figure}

To investigate the impact of monomer doping in the quantum dynamical behaviors, we calculate the dispersions of dynamical dimer density correlations within two scenarios: the strictly one-dimer-per-site manifold (without monomers) and manifolds doped with monomers at densities of 1.22\%, 25.0\%, and 49.8\%. 
These calculations involve sampling connected configurations and applying statistical averaging to approximate the system's dispersion,
and the result is shown in Fig.~\ref{fig:dimer-sma}. 
Without monomer doping, there exhibits a gapless mode with quadratic dispersion at $(\pi,\pi,\pi)$ that corresponds to the ``photon'' excitations of the U(1) Coulomb liquid. 
This photon mode is gapped out upon monomer doping, coincides with that observed in the pyrochlore lattice, and can be interpreted as the condensation of monomers that Higgses the U(1) gauge field, destroys the deconfined Coulomb phase and results a trivial confined phase.

%#####################################################

\begin{figure} [htbp]
\includegraphics[width=0.25\textwidth]{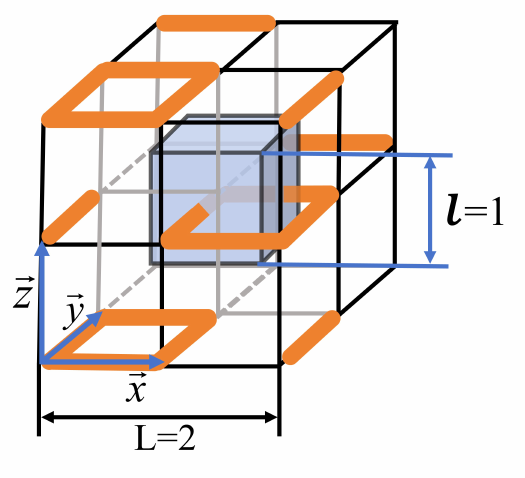}
\caption{\label{fig:string} Surface in cubic lattice. The model illustrates one of the configurations of a two-dimer per site within a dimer model of size $L=2$
that satisfies periodic boundary conditions. In the three-dimensional dimer model, the surface is a cube with edge length $(l=1)$, surrounding the lattice site. Some dimers are intersected perpendicularly by the surface.
The direct lattice vectors in the $\overrightarrow{x},\overrightarrow{y}$ and $\overrightarrow{z}$  directions are set to have a unit length of 1. }
\end{figure}

The fate of the Coulomb liquids upon doping can be also examined by probing the non-local ``surface'' operators, which serve as a diagnostic for quantum spin liquids (QSLs). 
Here, we focus on the diagonal operator $ Z =\prod_{i\in s} \sigma_i^z$, where $\prod_{i\in s}$ denotes the product of dimers that intersect with the surface $s$, $\sigma_i^z=1-2n_i$, $n_i=0$ or $1$ respectively indicates whether there is a dimer at position $i$, see Fig.~\ref{fig:string}.
This non-local operator $Z$, which measures the parity of the dimers intersecting with the surface, characterizes how well the local dimer constraint is satisfied. 
For example, for the smallest cubic surface that encloses only a single site of the cubic lattice (Fig. ~\ref{fig:string}), $\langle Z \rangle = \pm 1$ holds for any undoped dimer configurations, where the sign depends on the parity of the number of dimers attached to each site. 
For generic surfaces, $\langle Z \rangle = 1$ holds for configurations that each site touches even number of dimers, while  $\langle Z \rangle = (-1)^\textrm{\# of enclosed sites}$ for configurations that each site touches odd number of dimers. 
In presence of monomer doping, the local constraints are no longer strictly satisfied, leading to a decrease in $|\langle Z \rangle |$.
Also note that in two-dimensional lattices, such surfaces correspond to closed strings, hence the definition of $Z$ devolves into that in Ref. \cite{Semeghini21}. Due to the complexity of the diamond lattice structure, the actual calculation of this physical quantity is rather complicated. Considering that the physics is the same as in the cubic lattice, we have therefore calculated the surface operator only for the cubic lattice.
 
\begin{figure} [htbp]
\includegraphics[width=0.4\textwidth]{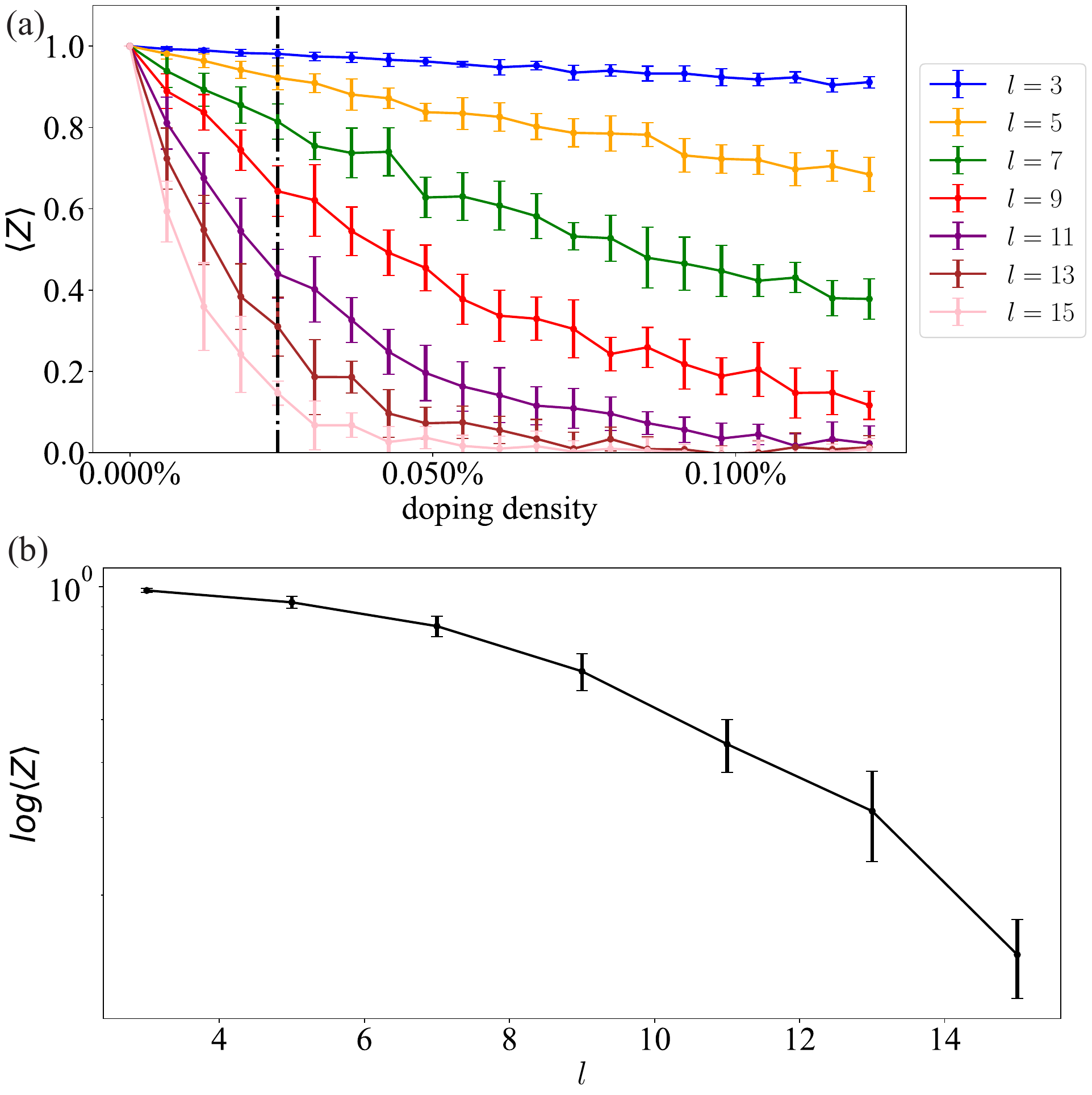}
\caption{\label{fig:3} (a) Surface operator as a function of the doping density.  (b) Surface operator as a function of the liner sizes %of string
for monomer density 0.0244\% (black dashed line in (a)).}
\end{figure}

We consider doping based on background configurations of two dimers per site. In Fig.~\ref{fig:3}(a), we plot the expectation value of the non-local operator $\langle Z \rangle $ versus the monomer density for different surface sizes with linear dimension $(l=3,5,7,9,11,13$ and $15)$. 
At the generalized RK point, the expectation value of the non-local surface operator $\langle Z\rangle$ can be calculated as $\langle Z \rangle=\frac{1}{N_C} \sum_{C} (-1)^ {m(C)}$,  where $(-1)^ {m(C)}$ measures the parity of the dimers intersecting with the surface for a given configuration $C$, and $N_C$ is the total number of configurations $C$ connected by local dimer moves in the doped QDM Eq. \eqref{eq:doped_QDM_cubic}.
As shown in Fig.~\ref{fig:3}(a), we find that 
with increasing monomer doping, $\langle Z \rangle $ rapidly decreases to $0$. In addition, with the increase as of the surface size $l$, $\langle Z \rangle $ also rapidly decreases.

To figure out whether infinitesimal amount of monomer defects could destroy the structure of the U(1) Coulomb liquid, we choose a configuration with $0.024414\% $ monomer doping on the background of two dimer per site, and examine the evolution of surface operator expectation value $\langle Z \rangle $ with increasing surface size $l$. 
As shown in Fig.~\ref{fig:3}(b), we find a superexponential decay of $\langle Z \rangle $ with increasing $l$, suggesting that the U(1) Coulomb liquid is unstable against infinitesimal monomers in the thermodynamic limit. 
Nevertheless, in cold-atom experiments where the atom arrays are prepared with limited system sizes, we still observe a pronounced $\langle Z \rangle $ when monomer doping is small. 
Hence we still expect to observe Coulomb liquid physics in cold-atom platform even if a small amount of monomer defects are presented. 
%----------------------------------------------------------

\begin{figure} [htbp]
\includegraphics[width=0.35\textwidth]{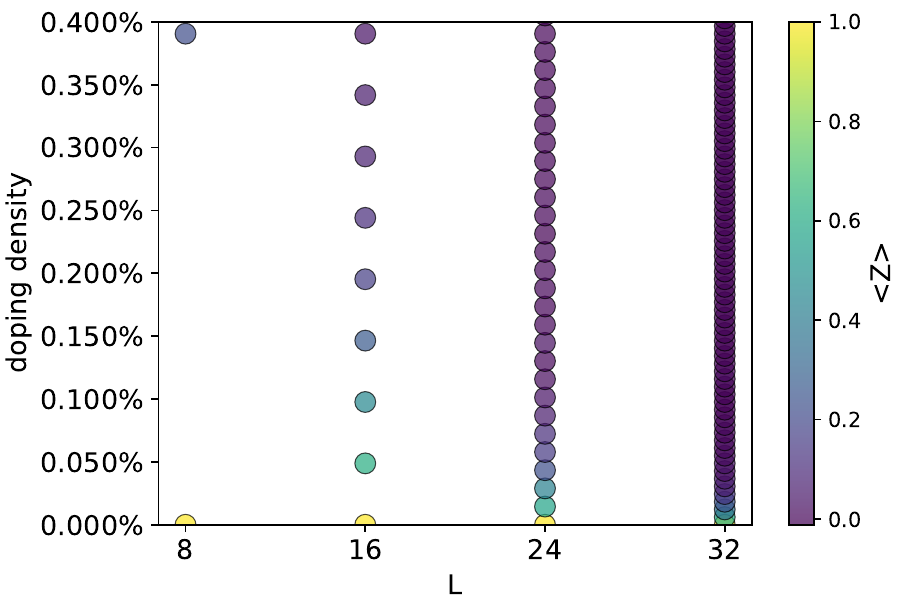}
\caption{\label{add} Stability phase diagram of the dimer model on the cubic lattice. The horizontal axis represents the doping density and the vertical axis denotes the system size ($L \times L \times L$). The color map indicates the expectation value of the surface operator. The surface size is $L/2-1$}
\end{figure}

\begin{figure} [htbp]
\includegraphics[width=0.4\textwidth]{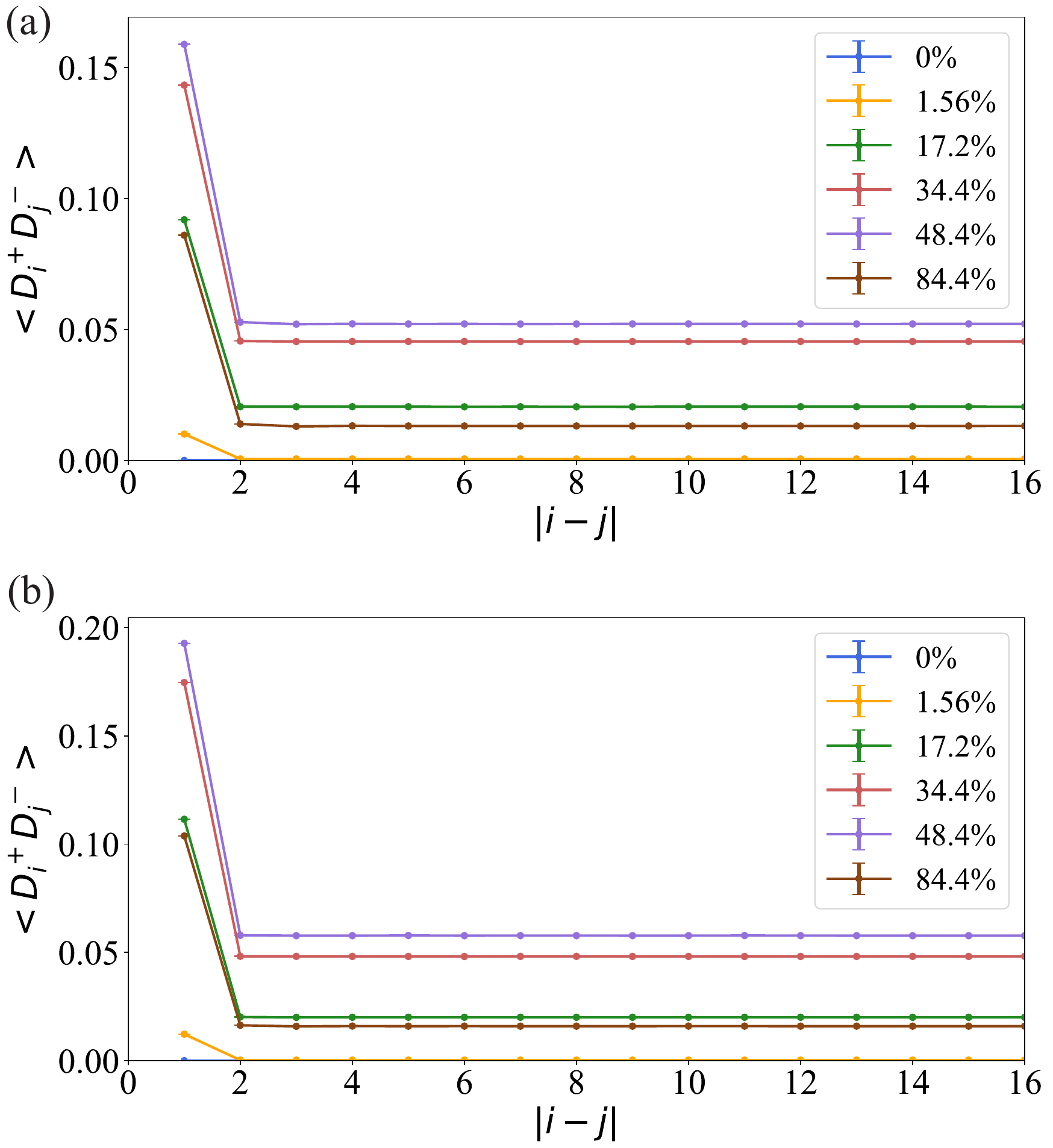}%
\caption{\label{fig:4} Dimer off-diagonal correlation at different doping density.
The circles represent the off-diagonal correlations along the $\overrightarrow{x}$  for the different distances $|i-j|$ in a cubic lattice. Different colors indicate varying doping densities ($ 0\%, 1.56\%, 17.2\%,34.4\%,48.4\%$ and $84.4\%$) in (a) one-dimer-per-site and (b) two-dimers-per-site configurations. The distance between two neighboring links is set to 1
} 
\end{figure}

To clearly illustrate the crossover between effective stable states in small system sizes, we present in Fig.~\ref{add} the “stability phase diagram” of the model. We calculate the expectation value of a three-dimensional surface operator $\langle Z \rangle$ of scale $l = L/2 - 1$ for systems of different sizes ($L \times L \times L$, $L=8,16,24,32$). Under the background of two dimers per site, the color map of $\langle Z \rangle$ is used to characterize the effective U(1) Coulomb liquid at different doping densities. As shown in Fig.~\ref{add}, the results are fully consistent with the preceding analysis: the characteristic signal of the U(1) Coulomb liquid is extremely sensitive to the size of the system. However, in finite systems this signal still emerges within a very narrow region of doping density.

We have also evaluated the dimer off-diagonal correlations to probe the possibility of proximate long-range order. 
The detailed calculation scheme is similar to that described in Sec. \ref{section:Pyrochlore lattice}, and the results are shown in Fig.~\ref{fig:4}. 
In the absence of monomer defects, we find vanishing off-diagonal dimer correlations as is forbidden by the local constraints, similar to the case of the diamond lattice.
As the local constraint becomes weakened through the introduction of monomers, 
the system immediately establishes a finite off-diagonal long-range correlation as a consequence of monomer condensation.

\begin{figure} [htbp]
\includegraphics[width=0.4\textwidth]{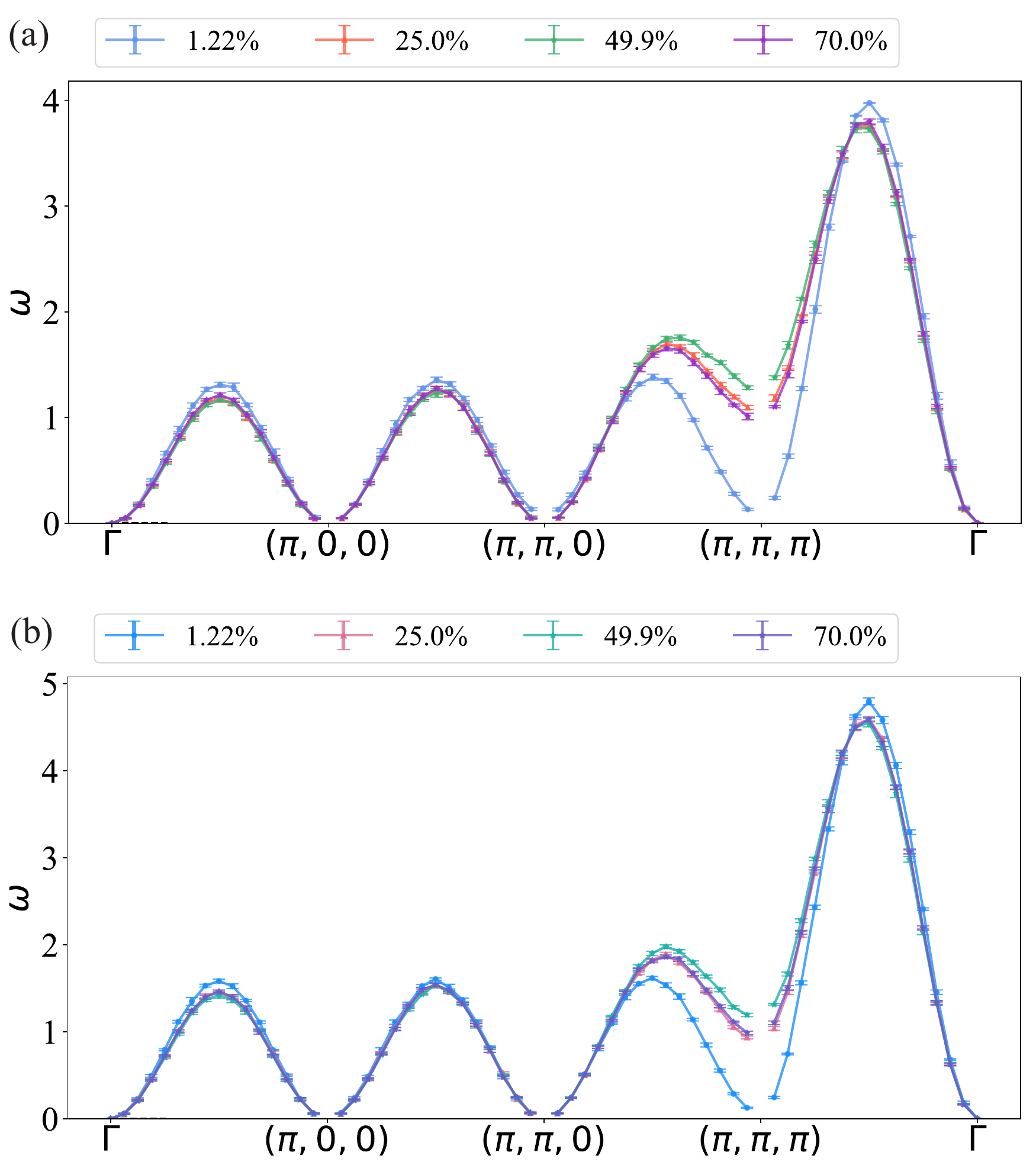}%
\caption{\label{fig:5} Dispersion of monomer density correlation along the paths $\Gamma \rightarrow (\pi,0,0) \rightarrow (\pi,\pi,0) \rightarrow (\pi,\pi,\pi) \rightarrow \Gamma$ for configurations based on the one dimer per site (a) and two dimers per site (b), with doping density of $1.22\%$,  $25.0\% $, 49.9\% and 70.0\%.}
\end{figure}

\begin{figure} [htbp]
\includegraphics[width=0.5\textwidth]{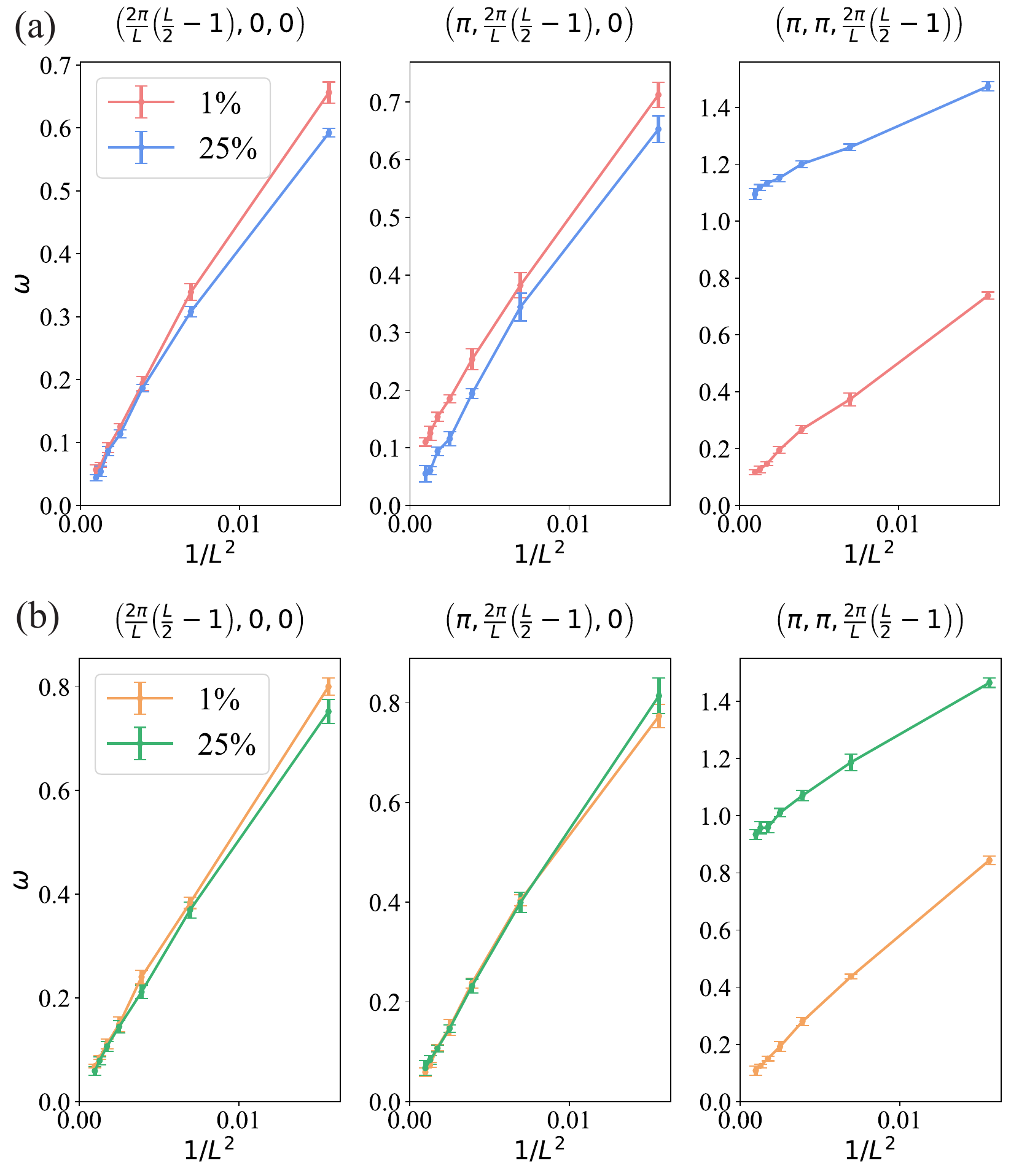}%
\caption{\label{fig:6} Finite size scaling of gap at ($\pi$,0,0), ($\pi$,$\pi$,0) and ($\pi$,$\pi$,$\pi$). The gap of monomers is shown as a function of the inverse squared system size for (a) one dimer per site and (b) two dimers per site doped with $1\%$ and $25\%$ monomers, respectively, at $\left(\frac{2\pi}{L}  \left(\frac{L}{2}-1 \right), 0, 0\right)$, $\left(\pi,\frac{2\pi}{L}  \left(\frac{L}{2}-1 \right), 0\right)$ and $(\pi,\pi, \frac{2\pi}{L}  (\frac{L}{2}-1) )$. }
\end{figure}

\begin{figure} [htbp]
\includegraphics[width=0.45\textwidth]{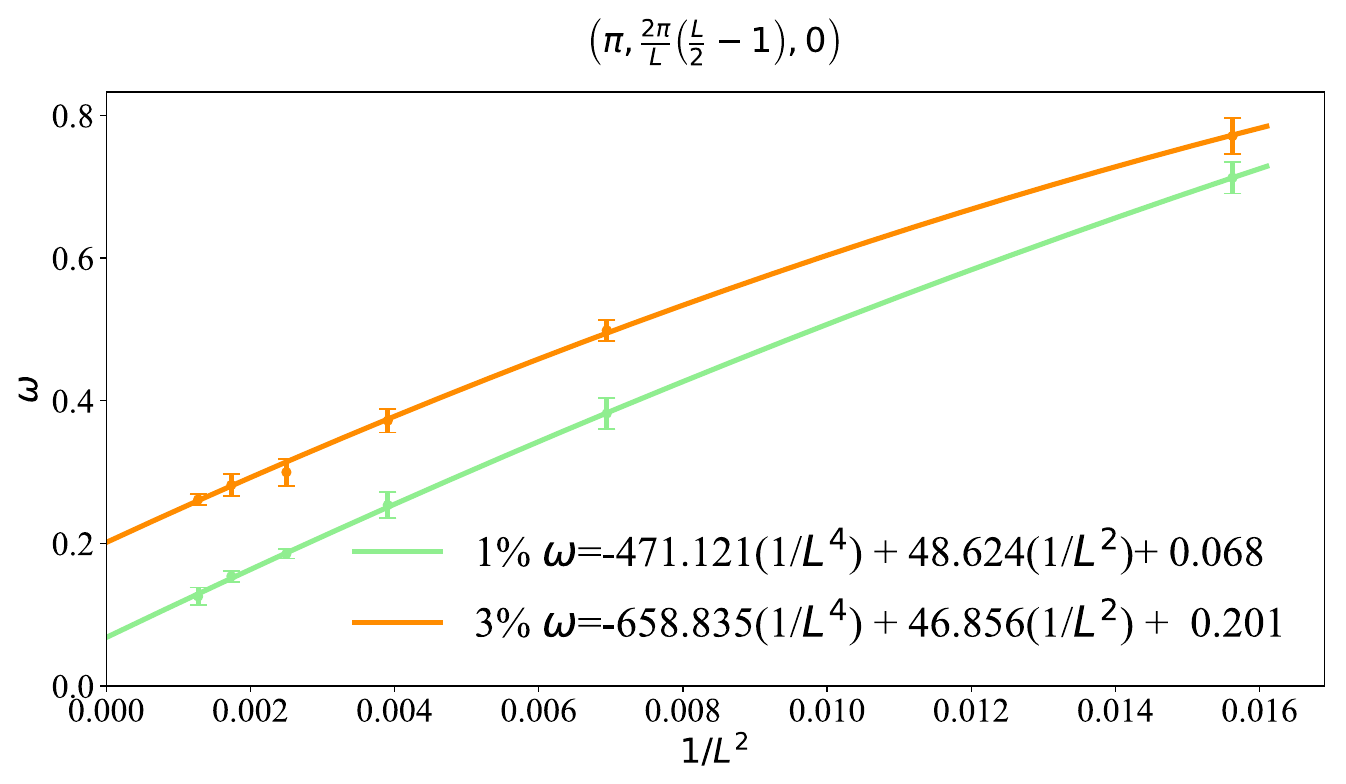}%
\caption{\label{fig:7} Fitting of the gap at $\left(\pi,\frac{2\pi}{L}  \left(\frac{L}{2}-1 \right), 0\right)$ under doping density 1\% and 3\%. After doping on one-dimer-per-site background, the gap is opened at $\left(\pi,\pi, 0\right)$.}
\end{figure}

We also evaluated the dispersion of monomer density correlations in the cubic lattice along the path $\Gamma \rightarrow \left(\pi,0,0\right) \rightarrow \left(\pi,\pi,0\right) \rightarrow \left(\pi,\pi,\pi\right) \rightarrow \Gamma$ based on method described in \ref{section:Pyrochlore lattice}. 
In  Fig.~\ref{fig:5}, we present the results based on one-dimer-per-site and two-dimers-per-site configurations with doping density ($1.22\%$, $25.0\%$, $49.9\%$ and $70.0\%$). In our SMA calculations, 
we observed that at the momentum points
 $\left(\pi,0,0\right)$,  $\left(\pi,\pi,0\right)$ and $ \left(\pi,\pi,\pi\right)$, both the numerator and the denominator (the structure factor) of $\omega_{SMA}$ approach zero, resulting in an indeterminate ’$0/0$’ form. This makes it impossible to directly determine whether $\omega_{SMA}$ vanishes at these points. 
To further investigate the nature of excitations at these points, we perform a finite-size scaling analysis, examining the asymptotic behavior of the dispersions at the nearby points $\left(\frac{2\pi}{L}  \left(\frac{L}{2}-1 \right), 0, 0\right)$, $\left(\pi,\frac{2\pi}{L}  \left(\frac{L}{2}-1 \right), 0\right)$ and $(\pi,\pi, \frac{2\pi}{L}  (\frac{L}{2}-1) )$ at different system sizes $L\times L\times L$. As $L \to \infty$, these adjacent momenta point approach the singular points.
As shown in Fig.~\ref{fig:6}, we find that the gap extrapolates to zero at the $\left(\pi,0,0\right)$ for both doped one-dimer-per-site and two-dimers-per-site states, and $\left(\pi,\pi,0\right)$ for only two-dimers-per-site state with increasing system size, almost independent of the monomer density. Fig.~\ref{fig:7} displays that the gap at $\left(\pi,\pi,0\right)$ for one-dimer-per-site state will open after doping.
In contrast, the dispersion at $ \left(\pi,\pi,\pi\right)$ extrapolates to some finite values and increases upon monomer doping.
Therefore, we conclude that in presence of monomer doping, excitations at the $\Gamma$, $\left(\pi,0,0\right)$ and $\left(\pi,\pi,0\right)$ points remain gapless while excitations at  $ \left(\pi,\pi,\pi\right)$ becomes gapped out. 

Note that the gapless mode at the $\Gamma$ point also appears in the diamond lattice system and is interpreted as the Goldstone mode above the monomer condensates. 
In contrast, the gapless excitations at $(\pi, 0, 0)$ and $(\pi, \pi, 0)$ are unique to the cubic lattice and must be attributed to other mechanisms. 
In fact, this scenario are closely parallel with the excitation spectra of the square and honeycomb lattice QDMs, where additional gapless ``pi0n'' excitations emerge alongside to the resonon (photon) excitations~\cite{moessner2008quantum}, and are attributed to the proximity of the long-range ordered valence bond phases.  
Therefore, the presence of these gapless modes implies that this generalized RK point does not lie as a part of an extended phase; instead, it corresponds to a quantum critical point proximate to various long-range ordered phases. 
Investigation of the extended phase diagram away from the generalized RK points involves methods beyond SMA such as quantum Monte Carlo~\cite{ZY2019sweeping,ZY2020improved,yan2019widely,dabholkar2022reentrance}, and is beyond the scope of the current work. 

\begin{comment}
%#############################################
\end{comment}

\section{Discussion and outlook}
%###################################################
Motivated by the recent developments on Rydberg atom arrays, 
in this work we investigate the relevant extended QDMs on three-dimensional bipartite lattice geometries with the introduction of monomers. 
For both diamond and cubic lattices,
we find that even a small amount of monomer doping disrupts the deconfined U(1) QSI
%Coulomb liquid 
and stabilizes off-diagonal long-range order. 
However, in a finite size (as in cold-atom experiment), the property of QSI will be kept in a certain region like a crossover after doping. 

Our results on the extended QDM have significant implications for experimental realizations, particularly in cold atom setups. 
One basic assumption in our model study is that monomer defects are condensed and enter the ground state wave functions upon doping. 
In practical experiments, monomers are often controlled by some effective chemical potential. 
If monomers remain gapped, they can be integrated out, leaving the U(1) Coulomb phase unaffected. 
However, if the monomer gap close and condensation occur, this would lead to the Higgsing of the U(1) gauge field, resulting in a trivial phase with off-diagonal long-range order in the thermodynamic limit. 
Despite this, real cold atom experiments are typically constrained by finite lattice sizes, which means that the system may not reach the thermodynamic limit. 
Consequently, we still expect the behavior of the U(1) Coulomb liquid to be in these experimental conditions if the doping is small. 

Our study of the extended QDM demonstrates a promising route to exploring the intricate interplay between charged matter and gauge fields in relevant physical systems. 
While the original RK-QDM completely ignores charge dynamics, our extended QDM explicitly includes monomers, allowing quantum dynamics of both the gauge field and monomer charges. 
This framework enables us to formulate a generalized version of the exactly solvable RK point, which can be efficiently tackled with classical Monte Carlo techniques. 
Moreover, away from this generalized RK point, the Hamiltonian remains sign-free, making it amenable to large-scale quantum Monte Carlo simulations. 
Consequently, our approach not only sheds light on the fundamental behavior of doped QDMs, but also paves the way for systematic studies of charged matter coupled to gauge fields that emerge in strongly correlated systems.

%###################################################
\begin{acknowledgments}
We thank the helpful discussions with Yang Qi, Fabien Alet, Sylvain Capponi and Pranay Patil. Y.C.W. and C.L acknowledges the support from the Natural Science Foundation of China (Grant No. 12474216 and 12564021) and Zhejiang Provincial Natural Science Foundation of China (Grant No. LZ23A040003), and the support from the High-Performance Computing Centre of Hangzhou International Innovation Institute of Beihang University. The work is supported by the the Scientific Research Project (No.WU2024B027) and start-up funding of Westlake University. The authors thank the high-performance computing center of Westlake University and Beijng PARATERA Tech Co.,Ltd. for providing HPC resources.
\end{acknowledgments}

%\appendix

%\nocite{*}

\bibliography{apssamp}% Produces the bibliography via BibTeX.

\end{document}